\documentclass[acmsmall,screen,nonacm]{acmart}
\AtBeginDocument{%
  }

\usepackage{soul}

\usepackage{csquotes}
\usepackage{natbib}
\usepackage{tikz}
\usetikzlibrary{shapes,arrows,decorations.pathmorphing,backgrounds,positioning,fit,automata}

\usepackage{enumitem}
\usepackage{xcolor}

\begin{document}

\title{``It Was a Magical Box'': Understanding Practitioner Workflows and Needs in Optimization}



\author{Connor Lawless}
\affiliation{%
  \institution{Stanford University}
  \city{Palo Alto}
  \state{California}
  \country{USA}
}
\email{lawlessc@stanford.edu}

\author{Jakob Schoeffer}
\affiliation{%
  \institution{University of Groningen}
  \city{Groningen}
  \country{The Netherlands}
}
\email{j.j.schoeffer@rug.nl}

\author{Madeleine Udell}
\affiliation{%
  \institution{Stanford University}
  \city{Palo Alto}
  \state{California}
  \country{USA}
}
\email{udell@stanford.edu}

\begin{abstract}
Optimization underpins decision-making in domains from healthcare to logistics, yet for many practitioners it remains a ``magical box'': powerful but opaque, difficult to use, and reliant on specialized expertise. While prior work has extensively studied machine learning workflows, the everyday practices of optimization model developers (OMDs) have received little attention. We conducted semi-structured interviews with 15 OMDs across diverse domains to examine how optimization is done in practice. Our findings reveal a highly iterative workflow spanning six stages: problem elicitation, data processing, model development, implementation, validation, and deployment. Importantly, we find that optimization practice is not only about algorithms that deliver better \textit{decisions}, but is equally shaped by \textit{data} and \textit{dialogue}---the ongoing communication with stakeholders that enables problem framing, calibrated trust, and adoption. We discuss opportunities for future tooling that foregrounds data and dialogue alongside decision-making, opening new directions for human-centered optimization.
\end{abstract}






\maketitle

\section{Introduction}

Optimization problems lie at the heart of some of today’s most pressing challenges: coordinating sustainable energy generation \cite{bazmi2011sustainable}, delivering healthcare efficiently \cite{bertsimas2024hospital}, and managing global supply chains \cite{bramel1998logic}.
At its core, optimization provides a mathematical framework for selecting the best option from a vast set of alternatives under different constraints: for example, minimizing delivery costs while ensuring that every customer is served, or allocating staff to balance workload and availability.
Importantly, optimization is not only central to these high-stakes domains; it also plays an increasingly prominent role within the fields of \textit{computer-supported cooperative work} (CSCW) and \textit{human-computer interaction} (HCI), powering systems for engineering design \cite{burnell2020gpkit}, design-parameter tuning \cite{niwa2025cooperative}, and preference elicitation \cite{lawless2024want}.
Decades of progress in algorithms and solvers, such as branch-and-bound algorithms for mixed-integer linear programming, have made it possible to solve large-scale problems with millions of variables and constraints \cite{wolsey2020integer}.
Yet, applying optimization in practice remains far from straightforward: it is a socio-technical process that requires translating ambiguous business needs into precise mathematical formulations.

Despite its promise, optimization remains accessible only to a narrow group of experts.
Commercial solver usage reflects this disparity: more than 80\% of Gurobi, a leading solver for optimization, users hold graduate degrees \cite{gurobiReport}, a signal of the steep expertise required to model and solve real-world optimization problems.
The optimization community has recently diagnosed why, contrasting optimization's trajectory with that of machine learning (ML), whose accessible tooling have enabled far wider participation \cite{le2025making, simchi2025democratizing}.
Some of the resulting barriers are not addressable through tool or interface design: the theory behind modern solvers (e.g., interior-point and branch-and-cut methods) is graduate-level material, and no amount of human-centered design substitutes for that expertise.
But others are squarely design problems, including limited practitioner-facing tooling and documentation, and, we argue, a specific communication gap.
As \citet{le2025making} observe, optimization models are ``inherently interpretable in theory'' (i.e., have an explicit objective, explicit constraints, and explicit variables should make a decision legible) yet ``in practice, the scale, complexity, and constraints of real-world problems make it difficult to communicate the rationale behind optimization outcomes to stakeholders,'' and this perceived opacity is what limits adoption.
This gap between formal interpretability and practical understandability is not simply a matter of explaining an algorithm’s output: stakeholders may also need to understand how their goals were translated into objectives and constraints, which assumptions shaped the model, and what trade-offs informed its recommendations.
These challenges are compounded when stakeholders and optimization practitioners have different forms of expertise and must work together to determine whether a model adequately reflects the problem at hand.
CSCW and HCI can contribute by studying and designing tools and practices that support this collaborative work.
For example, by helping practitioners elicit requirements, document modeling assumptions, communicate trade-offs, and engage stakeholders in validating solutions.
Such interventions cannot replace the specialized expertise required to formulate and solve complex optimization problems, but they can make the process through which optimization decisions are produced more transparent, collaborative, and responsive to stakeholder needs.
We return to this distinction in Section~\ref{sec:discussion}, where we examine which of these barriers our participants' accounts substantiate and how they compare to the accessibility challenges documented in ML.

In general, little is known about the workflows of optimization model developers (OMDs).
This gap lies in sharp contrast to ML, where practitioner workflows have been extensively documented through studies of data science practice (e.g.,~\cite{muller2019data, patel2008investigating,zhang2020data}), ML operations (e.g.,~\cite{shankar2024we,kreuzberger2023machine,matsui2022mlops}), and data documentation (e.g.,~\cite{heger2022understanding,bhat2023aspirations,winecoff2025improving}).
Unlike ML workflows that center on extracting insights from historical data to generate \textit{predictions}, optimization workflows revolve around generating actionable \textit{decisions}.
This distinction shapes practice: optimization is characterized by messy and incomplete data that inform and constrain model formulation, pragmatic trade-offs in computation and model fidelity, and sustained dialogue between different stakeholders that both shapes problem understanding and builds the appropriate trust needed for adoption.
Understanding these distinct workflows is essential for identifying bottlenecks and designing tools that make optimization more usable, trustworthy, and impactful.

To address this gap, we conducted semi-structured interviews with 15 OMDs working across manufacturing, logistics, healthcare, defense, and education.
OMDs recounted projects, diagrammed workflows, and reflected on pain points and opportunities.
From these accounts, we identified a highly iterative workflow consisting of problem elicitation, data processing, model formulation, implementation, validation, and deployment.
Beyond this workflow, our findings reveal how optimization practice is shaped not only by algorithms that generate better \textit{decisions} but also by the broader realities of \textit{data} and stakeholder \textit{dialogue} that make those decisions usable in practice.

This paper contributes a qualitative account of optimization practice that foregrounds the socio-technical realities of modeling work.
We offer \textit{(i)} an empirical characterization of how optimization workflows unfold across diverse domains, \textit{(ii)} a conceptual framing that highlights the overlooked roles of data and dialogue alongside decisions, and \textit{(iii)} design implications for tools and practices that better align with how optimization is carried out in the real world.
By surfacing these dimensions, our work provides a richer understanding of optimization practice and opens new directions for designing systems that can more effectively support it.
Our focus is on the barriers facing \textit{current} optimization practitioners, the population we study, rather than on democratizing optimization for non-experts from scratch.
We see this as a necessary first step rather than a substitute for that longer-term goal---tools that make expert practice more legible, efficient, and transferable are also the foundation on which broader access to optimization will eventually be built.

The remainder of this paper is organized as follows.
We begin by reviewing background work on optimization and related studies of practitioner workflows in Section~\ref{sec:related_work}.
We then describe our interview study, including participant recruitment, data collection, and analysis methods in Section~\ref{sec:methodology}.
In Section~\ref{sec:findings}, we present our findings on the six stages of the optimization workflow.
Section~\ref{sec:discussion} then elaborates on three cross-cutting themes that define successful optimization projects---\textit{data}, \textit{decisions}, and \textit{dialogue}---and considers their implications for tool design and future research.
We conclude by reflecting on the broader significance of studying optimization practice as a socio-technical endeavor.

\section{Background and Related Work} \label{sec:related_work}

This section reviews optimization as a decision-making framework (Section~\ref{rw:mathematical_optimization}) and surveys related work on practitioner tooling, interactive optimization, and emerging large language model (LLM)-based approaches (Section~\ref{rw:tooling}).
We then draw connections to research on AI workflows and practitioner needs (Section~\ref{rw:workflows}), underscoring how optimization remains relatively underexplored within CSCW and HCI.

\subsection{Mathematical Optimization}\label{rw:mathematical_optimization}

Optimization is a family of computational methods used to make complex decisions—such as planning delivery routes, allocating staff, scheduling manufacturing tasks, or configuring digital interfaces.
Optimization problems underpin the field of \textit{operations research}, which leverages optimization, among other tools, to improve management and decision-making \cite{winston2004operations}.
In all of these cases, people must choose a ``best'' set of actions while respecting real-world limits such as budgets, capacities, time windows, or physical constraints \cite{nocedal2006numerical}.
Because the number of possible choices can be enormous, organizations often rely on optimization tools to help identify good solutions efficiently.
Optimization is inherently model-driven: practitioners must translate high-level goals into precise mathematical formulations and constraints, often choosing among multiple representations with very different computational behavior.
In contrast, ML is largely data-driven, centering on model training and evaluation rather than constructing the problem structure itself.

\begin{figure}[t]
    \centering
    \includegraphics[width=0.9\linewidth]{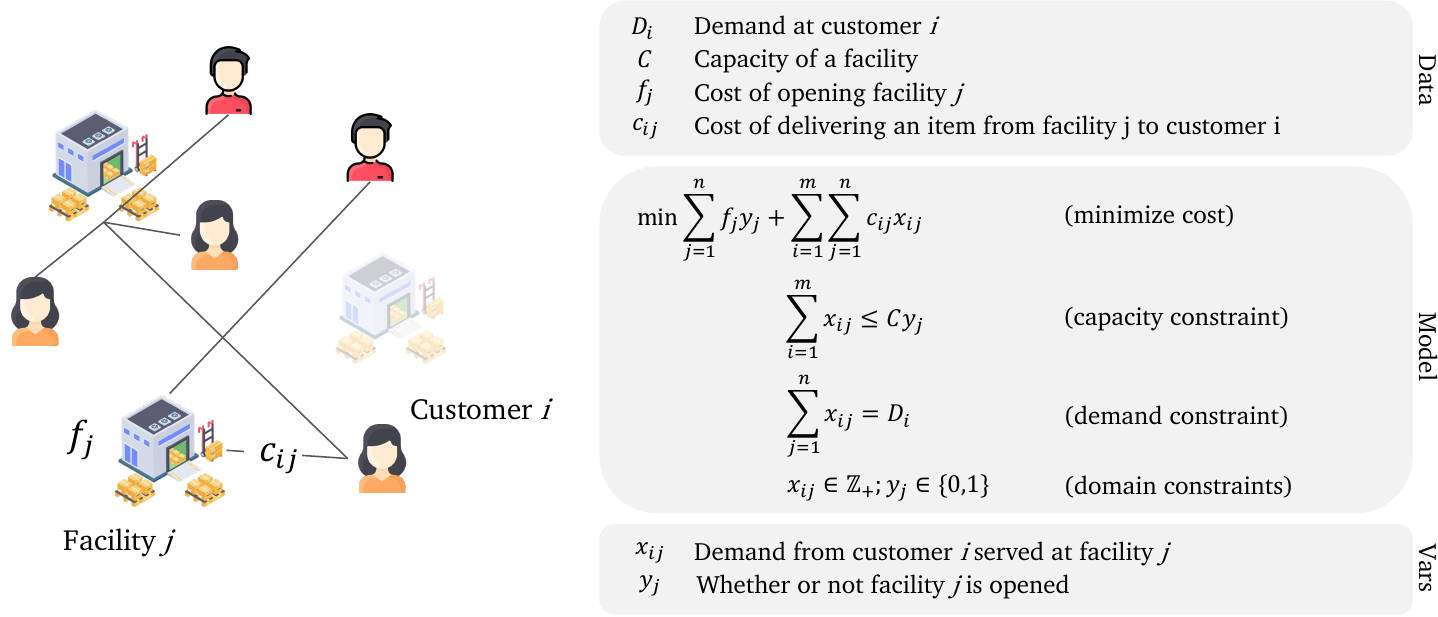}
    \Description{An example logistics application and its associated MILP model. (Left) A manufacturer is trying to decide what factories to build to serve a set of customers. The goal is to minimize the total cost of serving all the customers. (Right) Sample MILP model for the facility location problem. The model has two sets of decision variables: (i) $x_{ij}$ decides how much demand for customer $i$ is served by facility $j$, and (ii) $y_j$ decides whether to open facility $j$. The \textit{objective function} is to minimize the total cost which comprises both the cost of opening facilities and the cost of delivering items to meet the demand. The \textit{constraints} ensure that no demand is served by a facility is closed, all demand is served, and that no facility needs to serve more demand than it has capacity to meet.}
\caption{An example logistics application and its associated MILP model. (Left) A manufacturer decides what factories to build to serve a set of customers. The goal is to minimize the total cost of serving all the customers. (Right) Sample MILP model for the facility location problem. The model has two sets of decision variables: $(i)$ $x_{ij}$ decides how much demand for customer $i$ is served by facility $j$, and $(ii)$ $y_j$ decides whether to open facility $j$. The \textit{objective function} is to minimize the total cost which comprises both the cost of opening facilities and the cost of delivering items to meet the demand. The \textit{constraints} ensure that no demand is served by a facility that is closed, all demand is served, and that no facility needs to serve more demand than it has capacity to meet.}
    \label{fig:model}
\end{figure}

To solve these large-scale decision-making problems, practitioners translate informal requirements (e.g., ``every customer must get their package on time'') into a structured representation that can be solved using powerful commercial (e.g., Gurobi \cite{achterberg2019s}, CPLEX \cite{nickel2022decision}) and open-source (e.g., SCIP \cite{achterberg2009scip}) \textit{solvers}.
One of the most common representations is a \textit{mixed-integer linear program} (MILP) \cite{wolsey2020integer}:
\begin{itemize}
    \item \textit{Decision variables} represent choices (e.g., ``1 if truck A serves customer B, 0 otherwise'').
    \item \textit{Constraints} specify which combinations are allowed (e.g., ``a truck cannot exceed its capacity'').
    \item An \textit{objective function} defines what it means to do well (e.g., ``minimize total distance traveled'').
\end{itemize}
Although MILPs use mathematical notation, the underlying idea is intuitive: specify the choices, specify the rules, and let the solver search for the best valid combination.
See Figure \ref{fig:model} for an illustrative logistics example.

Importantly, there are often many different ways to represent the same real-world problem, depending on how practitioners choose to define decision variables or express constraints.
These alternative formulations can behave very differently when sent to a solver: some allow the solver to reason efficiently, while others lead to long runtimes or fail to find solutions at all \cite{IPref}.
For example, two equivalent formulations of the same delivery problem can differ by orders of magnitude in solver difficulty.
As a result, choosing a good formulation is a core part of practitioner expertise and a major source of iteration in real-world projects.
Scale introduces further complexity: many industrial problems are too large to solve exactly, and practitioners use \textit{decomposition algorithms} \cite{rahmaniani2017benders, desaulniers2006column, vanderbeck2006generic}---which break a large problem into coordinated subproblems---or \textit{heuristics} \cite{fischetti2010heuristics}, faster methods that trade guaranteed optimality for tractability.

\subsection{Tooling for Optimization Practitioners}\label{rw:tooling}

Traditionally, optimization tooling has centered on modeling languages that make it easier to express problems in solver-ready form.
Systems such as AMPL \cite{fourer1990ampl}, Convex.jl \cite{udell2014convex}, JuMP \cite{dunning2017jump}, CVXPY \cite{diamond2016cvxpy}, and solver-specific APIs like gurobipy~\citep{gurobipy} provide higher-level interfaces to connect with solvers.
Within the HCI community, GPKit introduced a human-centered Python package for geometric
programming, demonstrating how tool design can prioritize usability in addition to functionality \cite{burnell2020gpkit}.
These tools lower the barrier to model implementation but address only a narrow part of the workflow: translating mathematical formulations into solver-ready inputs.
In contrast, our study examines end-to-end workflows to identify opportunities for tooling beyond implementation.
Another line of work examines interactive optimization, where humans guide solver behavior (see \cite{meignan2015review} for a detailed review).
While these efforts emphasize human-in-the-loop interactions with structured elements of a solver, our work shifts attention to the broader workflows of optimization practitioners.

Most recently, advances in LLMs and natural language processing have motivated ``optimization copilots'' that aim to automate modeling end-to-end \cite{wasserkrug2024large, ramamonjison2023nl4opt,xiao24,ahmaditeshnizi2024optimus,huang2025orlm,astorga2025autoformulationmathematicaloptimizationmodels,huang2025orlm}.
Frameworks such as Chain-of-Experts \cite{xiao24} and OptiMUS \cite{ahmaditeshnizioptimus, ahmaditeshnizi2024optimus} leverage structured pipelines of LLM agents to model and solve MILP problems, while fine-tuned LLMs have been shown to improve formulation accuracy \cite{huang2025orlm}.
Building on these techniques, LLM-powered systems have been deployed in domains including supply chain optimization \cite{li2023large}, scheduling \cite{lawless2024want}, debugging \cite{chen2025optichat}, evaluating formulation equivalence \cite{zhaiequivamap}, and solver configuration \cite{lawless2025llms}.
However, most are designed as end-to-end solutions for non-experts, aiming to bypass rather than support practitioner expertise.
In contrast, our study examines how optimization experts actually work, surfacing workflow bottlenecks and opportunities for tools that augment rather than replace their expertise. 

Our framing of optimization's accessibility gap draws on \citet{le2025making}, who diagnose why optimization has not seen ML-scale adoption despite comparable technical power.
They point to a fragmented, largely proprietary tooling ecosystem, thin practitioner-facing pedagogical resources, and, notably, a mismatch between optimization's inherent interpretability and its perceived opacity to stakeholders.
\citet{le2025making} wrote a field-level position paper, and its proposed remedies are institutional and educational (e.g., curricula, funding, outreach) rather than design interventions.
Our study complements this diagnosis empirically: we examine which of these barriers manifest in day-to-day OMD practice, and which are addressable through the kind of tool and interface design CSCW and HCI can offer (Section~\ref{sec:discussion}).
In particular, the interpretability mismatch \citet{le2025making} identify is one of several concrete differences from ML practice that we ground in our findings and unpack in Section~\ref{sec:ml_comparison}.

\subsection{AI Practitioner Workflows and Needs}\label{rw:workflows}

CSCW and HCI have a long tradition of studying situated work practice: how people actually coordinate, communicate, and get things done, often in tension with how that work is formally represented or proceduralized~\citep{schmidt1992taking,suchman1987plans}.
A central contribution of this tradition has been to surface the socio-technical
processes and practices through which work gets done, many of which remain invisible in formal
workflow descriptions and the systems built around them.
Studies of print shop workers, for example, show how formal workflow systems are continuously reconciled with the improvisational practices of the shop floor~\citep{bowers1995workflow}.
Studies of control room operators show how real-time, safety-critical decisions are coordinated through informal, situated practices that resist full proceduralization~\citep{heath1992collaboration}.
Studies of clinical practice show how different healthcare practitioners coordinate around a single patient in light of different motivations and concerns~\citep{reddy2001coordinating}.
Across these settings, this research has made visible the often-invisible work of articulation, improvisation, and mutual adjustment that sustains professional and technical workflows.

More recently, this same lens has been extended to AI practitioners specifically, documenting how they navigate data work (e.g.,~\cite{muller2019data,sambasivan2021everyone}), model development and deployment (e.g.,~\cite{wang2019human,xin2021whither,shankar2024we}), collaboration and trust (e.g.,~\cite{passi2018trust,zhang2020data}), and documentation practices (e.g.,~\cite{bhat2023aspirations,heger2022understanding,winecoff2025improving}).
We draw on this broader literature because it shows that AI practitioners work through iterative, collaborative, and organizationally embedded processes~\citep{muller2019data,passi2018trust,zhang2020data}---dynamics that, as we show in Section~\ref{sec:findings}, extend to OMDs as well.
At the same time, most prior work has focused on prediction-oriented workflows in data science and ML rather than on the development of formal decision models.
We therefore use this literature as an adjacent point of comparison: it highlights important shared concerns while underscoring that optimization remains a distinct and relatively unstudied workflow domain within CSCW and HCI.

Ethnographic and interview studies show that data science is inherently collaborative and iterative~\citep{muller2019data,wang2019human}.
This work relies on an ``outer loop'' of dialogue, including building trust, framing problems, bridging gaps between technical and domain expertise, and helping clients process results~\citep{kross2021orienting,passi2018trust}.
Trustworthiness emerges through ongoing negotiation and the balancing of metrics with organizational intuition~\citep{passi2018trust}.
Like data scientists, optimization practitioners work through sustained dialogue with stakeholders (see Section~\ref{sec:dialogue}).
But optimization differs in a key respect: practitioners must not only interpret results, but also translate ambiguous organizational goals into explicit objectives, constraints, and decision variables~\citep{IPref,wolsey2020integer}.

Practitioners view automation as inevitable and useful for generating quick baselines and saving time, but not as a replacement for human judgment~\citep{wang2019human,xin2021whither}.
Instead, it supports a symbiotic workflow in which humans guide systems through defining search spaces and contributing domain knowledge~\citep{xin2021whither}.
Yet, prior work shows that tool adoption often remains pragmatic and improvisational: for example, fairness toolkits tend to fail when general functionality conflicts with context-specific definitions of fairness~\citep{deng2022exploring,passi2018trust}.
In creative domains, systems using \textit{intent tags} enable non-linear, iterative co-creation by giving users granular, flexible control across different abstraction levels---which can result in increased control and satisfaction relative to chat-based systems~\citep{gmeiner2025intent}.
In manufacturing, producing high-quality custom outputs with digital fabrication frequently requires practitioners to bypass off-the-shelf tools and instead write custom toolpath generators or directly edit low-level machine instructions to meet aesthetic or production constraints~\citep{hirsch2023nothing}.
Similarly, supporting complex multivariate statistical analysis demands tools that reduce error-detection latency and automatically propagate edits across specifications to manage cognitive load and deadlines~\citep{gu2023understanding}.
As we show in this work, optimization shares the reliance on human judgment (Sections~\ref{sec:lit-expert-judgment} and~\ref{sec:sniff-tests}) and workaround-heavy practice (Sections~\ref{sec:off-the-shelf} and~\ref{sec:portability}), but it adds a distinct modeling burden: practitioners must move among natural-language requirements and mathematical formulations (Section~\ref{sec:translate-math}), data pipelines (Section~\ref{sec:data_processing}), and solver implementations (Section~\ref{sec:model_implementation}), often comparing multiple valid formulations with very different computational behavior (Sections~\ref{sec:simplification-realism} and~\ref{sec:validate-scalability}).

Practitioners also face challenges in documenting their work and organizational practices.
Studies of documentation reveal tensions between transparency ideals (e.g., model cards~\citep{mitchell2019model}) and workplace realities.
\citet{heger2022understanding} found data documentation practices to be ad hoc, myopic, and driven by immediate utility rather than responsible AI reflection.
\citet{bhat2023aspirations} similarly identified a gap between proposed documentation frameworks and actual practice, particularly around ethical information.
Synthesizing this literature, \citet{winecoff2025improving} show that documentation’s intended role in governance is often undermined by insufficient incentives, limited resources, and weak integration with existing workflows.

This concern echoes a broader literature on data work in high-stakes AI.
\citet{sambasivan2021everyone} introduce the concept of \textit{data cascades}---compounding negative downstream effects that originate from technical debt or underinvestment in data quality, and that often surface far from their point of origin, which makes them difficult to trace and correct.
They find that data work is chronically undervalued relative to model-centric work across ML practitioners in high-stakes domains, echoing the lack of recognition afforded to data-centric labor documented elsewhere~\citep{bhat2023aspirations,heger2022understanding}.
Our findings extend this concern to optimization (see, e.g., Sections~\ref{sec:descriptions-vague} and~\ref{sec:data-problem}): because messy or incomplete data can silently distort problem formulation, constraint definitions, and validation, data issues in OMD workflows carry similar cascading risk---but with a twist, since they can invalidate the problem framing itself rather than only degrading downstream model performance.
OMDs likewise rely on documentation to coordinate work across teams (see Section~\ref{sec:doc-key}).
However, optimization documentation often centers on assumptions, constraints, heuristics, data definitions, and solver limitations needed to make a decision model usable in practice, not just transparent or accountable.

The rise of MLOps formalizes lessons from software engineering and DevOps into processes for deploying, monitoring, and maintaining ML systems~\citep{kreuzberger2023machine,matsui2022mlops,cois2014modern}.
Likewise, frameworks like CRISP-DM~\citep{wirth2000crisp} and CRISP-ML~\citep{studer2021towards} articulate idealized end-to-end pipelines that foreground the tight coupling between technical development, organizational processes, infrastructure, and governance~\citep{saltz2021crisp}.
Recent \textit{empirical} studies of MLOps, however, emphasize the sustained human effort required to bridge development and deployment.
For example, \citet{shankar2024we} characterize an iterative, human-centered MLOps workflow structured around managing \textit{velocity}, \textit{visibility}, and \textit{versioning}.
They show how engineers frequently build ``pipeline jungles''---hybrid mixtures of models and manual rules---as reactive fixes to maintain production stability.

Our work extends these insights to an overlooked community: optimization model developers.
Like ML practitioners, they balance technical rigor with organizational demands, but their challenges differ---translating domain expertise into formal constraints, navigating trade-offs between tractability and fidelity, and iteratively refining formulations as requirements evolve.
By examining their workflows, we broaden human-centered AI to include optimization, surfacing both shared concerns across AI practitioner communities and optimization’s unique requirements.
\section{Methodology} \label{sec:methodology}

The goal of this study was to better understand how OMDs approach their work, including the workflows they follow, the challenges they face, and the strategies they use to address them.
To this end, we conducted semi-structured interviews with 15 OMDs working across diverse domains.

\subsection{Study Participants}

We recruited participants via personal connections (N=6), advertising on professional networks and social media websites (N=2), advertising the study during a talk at a technical conference (N=4), and via snow-ball sampling with earlier participants (N=3).
Participants were required to complete a screening survey to ensure that they were over 18, based in the US, had obtained at least a bachelor's degree, and had completed at least one real-world project involving optimization.
The latter criterion was designed to ensure that participants had practical experience with optimization problems outside of an academic setting.

Since, to the best of our knowledge, there are no prior works that study OMDs, we focused on recruiting users with different education backgrounds and diverse application domains.
Table~\ref{tab:participants} summarizes the 15 participants of our study.
Of the 15 participants, 10 were male ($66\%$) and 5 were female ($33\%$).
8 of the participants ($53\%$) had obtained their bachelors, 4 ($27\%$) had a masters degree, and 3 had doctoral degrees ($20\%$) at the time of the interview.
Participants had, on average, 10 years of experience with optimization and worked on a number of applications including call center operations, health care delivery, logistics, manufacturing, and agriculture. 

\begin{table}[t]
    \small
    \centering
    \caption{Study participants.}
    \label{tab:participants}
    \begin{tabular}{l l l c l}
    \toprule
    \textbf{ID} &  \textbf{Role}  & \textbf{Highest Degree} & \textbf{Experience} & \textbf{Application Area}  \\
    \midrule
    P1 & Data Scientist  & Bachelors & 3--5 & Manufacturing \\
    P2 & PhD Student  & Masters & 6--10 & Agriculture \\
    P3 & PhD Student & Bachelors & 11+ & Healthcare \\
    P4 & Applied Scientist & Masters & 3--5 & Service Operations\\
    P5 & Algorithm Development Engineer & Masters & 3--5 & Defense\\
    P6 & Technical Fellow & PhD & 11+ & Logistics \\
    P7 & Operations Research Analyst & PhD & 11+ & Transportation \\
    P8 & PhD Student & Masters & 1--2 & Logistics \\
    P9 & Software Engineer & Bachelors & 1--2 & Education \\
    P10 & Optimization Consultant & PhD & 11+ & Consulting \\
    P11 & PhD Student & Masters & 3--5 & Logistics \\
    P12 & PhD Student & Bachelors & 3--5 & Education \\
    P13 & Applied Mathematician & Masters & 6--10 & Transportation \\
    P14 & PhD Student & Bachelors & 3--5 & Logistics \\
    P15 & Applied Scientist & Masters & 1--2 & Logistics \\
    \bottomrule
    \end{tabular}
\end{table}

\subsection{Interview Protocol}

Each participant attended a semi-structured interview conducted in-person (N=2) or virtually (N=13) via a video conference tool lasting between 35 to 75 minutes (average of 44 minutes).
All participants were compensated with a \$20 digital gift card.
During the interview, participants were asked to recall 1--2 prior projects that included the application of optimization to solve a real-world use case.
They then described the workflow of the project, from conception to deployment, including how they interfaced with different stakeholders and what tooling they used.
Participants diagrammed the process while describing the project to encourage reflection on the entire workflow.
Participants were asked to highlight pain points and bottlenecks in the project workflow and to brainstorm tools that could have helped improve their process.
After discussing the first project, if time permitted, participants were asked to describe a second project by contrasting it with the first. 
The interview guide can be found in Appendix \ref{app:interview_protocol}.
All interviews were captured via audio recordings.
The protocol was approved by an institutional review board at Stanford University.

\subsection{Coding and Analysis}

In total, 654 minutes of interviews were audio-recorded and transcribed using Marvin\footnote{https://app.heymarvin.com/}.
We analyzed the interview transcripts using an inductive reflexive thematic analysis approach \cite{braun2012thematic,clarke2017thematic}.
To ensure close engagement with the data and capture participants' perspectives, each interview transcript was initially open-coded independently by one of the researchers on a line-by-line basis.
This stage drew on open-coding practices from grounded theory \cite{charmaz2006constructing}.
These initial codes were then merged and grouped together to identify key patterns and relationships between different concepts.
After that, the team distilled these codes into 17 core themes that represent the key findings of the study.
Finally, the 17 core themes were systematically grouped and mapped onto a set of distinct yet interrelated workflow steps.
Importantly, these steps were not predefined but instead emerged inductively from the codes and themes.
An abridged version of the final codebook is included in Appendix \ref{app:codebook}.
Throughout the analysis, the researchers met weekly to discuss the emerging codes, resolve discrepancies, and iteratively reach a consensus on overarching themes and workflow steps. 
We did not calculate inter-rater reliability to avoid potential marginalization or minimization of perspectives~\citep{mcdonald2019reliability}.

\paragraph{Positionality}
The research team has prior training and professional experience in optimization and HCI, which positions us as partial insiders to the practices we study.
All authors have formal training in optimization, and the interviews were conducted by a researcher with a background in optimization, which facilitated rapport and allowed participants to discuss technical modeling decisions without extensive explanation.
At the same time, this shared expertise shaped the study in important ways.
Participants often assumed common ground, used domain shorthand, and foregrounded modeling considerations they viewed as salient within optimization communities.
During analysis, our familiarity with optimization workflows risked normalizing certain practices (e.g., informal validation strategies or solver-centric framings) rather than questioning them.
To mitigate this, we engaged in reflexive discussions across authors, revisited moments of assumed understanding in transcripts, and iteratively questioned interpretations that aligned too closely with our own modeling intuitions.
Nevertheless, our findings reflect perspectives produced through researchers embedded in optimization cultures, and other viewpoints may surface---to some degree---different concerns or interpretations.
\section{Findings} \label{sec:findings}

\begin{figure}[t]
    \centering
    \includegraphics[width=1\linewidth]{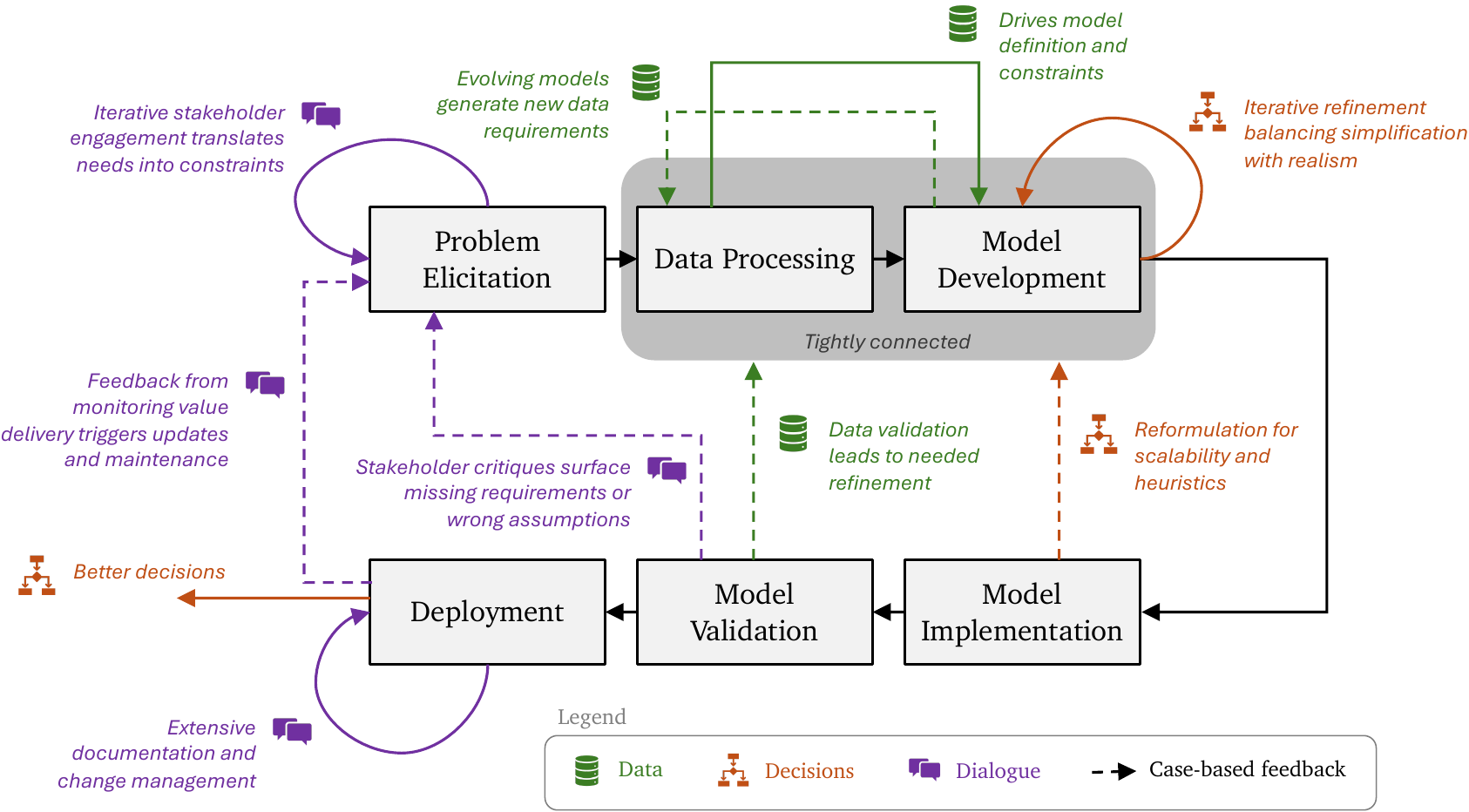}
    \Description{A six-stage workflow of optimization model development. The stages are: Problem Elicitation, Data Processing, Model Development, Model Implementation, Model Validation, and Deployment. The workflow is highly iterative with frequent backtracking and refinement across stages.}
    \caption{Six-stage workflow of optimization model development. The process spans problem elicitation, data processing, model development, model implementation, model validation, and deployment. Practitioners described this workflow as highly iterative, with frequent backtracking and refinement across stages.}
    \label{fig:workflow}
\end{figure}

Our interviews revealed that optimization model development is not a straightforward technical pipeline but an \textit{iterative socio-technical} workflow.
Practitioners consistently described moving back and forth across six interconnected stages: problem elicitation, data processing, model development, implementation, validation, and deployment (see Figure \ref{fig:workflow}).
Although each stage had its own bottlenecks and strategies, collectively they constituted a cycle of refinement where iteration was driven by changing requirements, imperfect data, and sustained dialogue with stakeholders.
To present these findings, we structure the remainder of this section around the six stages of the workflow, illustrating how practitioners approach each stage, the challenges they encounter, and the strategies they adopt to overcome them.
Although the stages may look familiar compared to ML development \cite{wang2019human, shankar2024we}, they differ significantly in what they entail.
Table~\ref{tab:summary} summarizes our findings along the optimization workflow.

\begin{table}[t]
    \footnotesize
    \centering
    \caption{Summary of optimization workflow insights and challenges.}
    \label{tab:summary}
    \resizebox{\textwidth}{!}{
    \begin{tabular}{p{2.3cm} p{4.0cm} p{7.7cm}}
        \toprule
         \textbf{Workflow Step} & \textbf{Insight} & \textbf{Exemplary Quote} \\
         \midrule
         Problem Elicitation & Initial problem descriptions are vague, requiring iterative stakeholder engagement & \textit{``If I had [...] just worked in my office and never had that dialogue [with stakeholders], the project would have never really gotten off the ground.'' --- P6} \\
         \cmidrule{2-3}
         & Problem understanding requires reconciling distributed expertise & \textit{``With this project in particular it was the sheer number of stakeholders that we had to communicate with that I think was especially challenging.'' --- P9} \\
         \cmidrule{2-3}
         & OMDs critically assess when optimization is (and is not) appropriate & \textit{``Sometimes the problem they may think is optimization, sometimes it's just about business process change.'' --- P10} \\
         \midrule
         Data Processing & Data tends to be messy, sparse, or incomplete & \textit{``Their data is not in a single place, or it's like, layered on top of a bunch of legacy systems.'' --- P1} \\
         \cmidrule{2-3}
         & Accessing data is slow and difficult & \textit{``It's not like from textbook [...] you really need to figure out whether you can get exactly the data you want.'' --- P3}  \\
         \cmidrule{2-3}
         & Data drives problem definition and modeling & \textit{``A lot of times the data will drive the knowledge, the understanding of the problem, frankly.'' --- P7} \\
         \midrule
         Model Development & Modeling is iterative and evolves with shifting requirements & \textit{``We'll go to formulate, come up with a solution, then put it out there, gather feedback and literally come back to [...] what's really your goal?'' ---~P4} \\
         \cmidrule{2-3}
         & OMDs balance simplification with realism & \textit{``If [...] it's more of a real-time application and we have millions of variables [...] we're going to have to go to a heuristic right away.'' --- P7} \\
         \cmidrule{2-3}
         & OMDs leverage literature and expert judgment & \textit{``I would build off of the ones that I found [in literature] but alter it based on the type of work we were trying to do.'' --- P8} \\
         \midrule
         Model Implementation & Implementation advances through iterative prototyping & \textit{``I go back and what I'll do is like run something with no constraints and then iteratively add constraints [...] does the solution behave like I expect it would?'' --- P8}  \\
         \cmidrule{2-3}
         & Initial implementations validate the scalability of the proposed model & \textit{``We started out with a single model and the runtime would be like over 24 hours or so, which we, like, definitely could not do.'' --- P12}  \\
         \cmidrule{2-3}
          & OMDs struggle to leverage off-the-shelf optimization solvers & \textit{``So really the only thing to do is to kind of just tweak the, the parameters and, and see what happens.'' --- P15} \\
         \midrule
         Model Validation & Models are validated through inspection and stakeholder critique & \textit{``[We] evaluate the solution using some common sense and then go back to the customer, discuss the model, gather some feedback and go to step one'' --- P4}  \\
         \cmidrule{2-3} 
          & OMDs satisfice with respect to computation times & \textit{``We do have some requests from them to about how long this algorithm should take'' --- P14}  \\
         \cmidrule{2-3}
          & Model validation is concurrent with data validation & \textit{``What combo data and constraints is causing a solution that's non intuitive.'' --- P7} \\
         \midrule
         Deployment & Documentation is key to successful deployments & \textit{``I would say that writing the document itself takes, takes some time'' --- P4}  \\
         \cmidrule{2-3} 
         & Portability and integration constraints shape deployment & \textit{``So we learned pretty late on that the third party didn't support what we needed them to support to be able to display the outputs of our model.'' ---~P1}  \\
         \bottomrule
    \end{tabular}
    }
\end{table}

\subsection{Problem Elicitation}\label{sec:problem-elicitation}

\begin{figure}[t]
    \centering
    \centerline{
    \includegraphics[width=\textwidth]{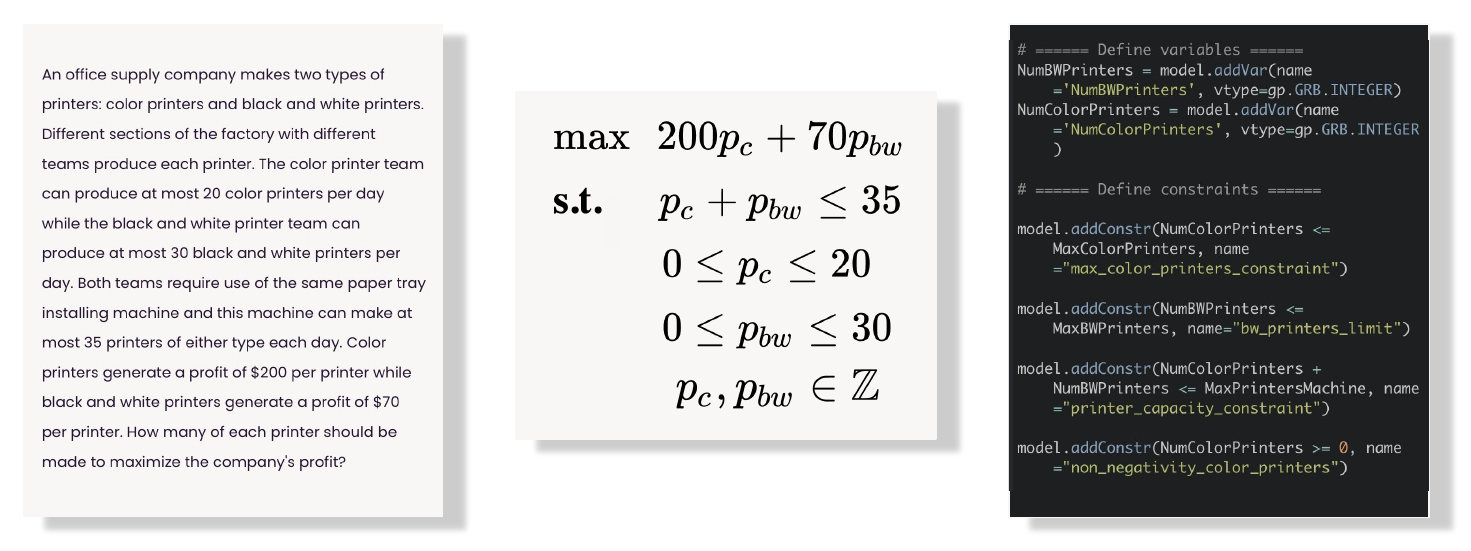}
    }
    \Description{Three sample artifacts illustrate stages of the optimization workflow. On the left, a natural language problem description explains a company’s printer production constraints. In the center, a formal mathematical optimization model expresses the same problem with an objective function and constraints. On the right, a Python code snippet using gurobipy~\citep{gurobipy} shows the model implemented programmatically.}
    \caption{Sample artifacts from along the optimization workflow. (Left) A sample document outlining the \textit{problem requirements} in natural language. (Center) A formal \textit{optimization model} that translates the problem description into mathematics. (Right) A sample \textit{implementation} of the model in Python using gurobipy~\citep{gurobipy}.}
    \label{fig:artifacts}
\end{figure}

\begin{displayquote}
    \textit{``The whole thing was all about this dialogue and interaction between me and the [stakeholders]. If I had kind of gone off and just worked in my office and never had that dialogue, the project would have never really gotten off the ground.'' \\ --- P6}
\end{displayquote}
The problem elicitation stage of the optimization workflow engages stakeholders to understand, define, and refine the problem optimization should address.
It translates high-level, often vague business needs into clear, structured requirements guiding model development (P3, P6, P12).
This involves assessing the client's current situation, determining if optimization is appropriate, and aligning diverse stakeholders on the problem (P10).
Since knowledge is distributed across business, technical, and customer experts (P4, P9), elicitation relies on iterative discussions, careful questioning, and non-technical communication (P3--6, P14--15).
A typical artifact is a \textit{problem requirements document} (P4, P10) detailing the decision-making problem in natural language (see Figure \ref{fig:artifacts}).
Although time- and resource-intensive, OMDs stressed that a thorough understanding of the problem underpins any successful project.
Prior data science-focused literature frames this kind of engagement as an ``outer loop'' that recurs throughout a project rather than a discrete upfront step~\citep{kross2021orienting}.
Our findings confirm that pattern, but also surface something the outer-loop framing does not capture on its own: elicitation in optimization must additionally produce a precise decision space, not just a shared understanding of the problem (see also Section~\ref{sec:ml_comparison}).

\subsubsection{Initial problem descriptions are vague, requiring iterative stakeholder engagement}\label{sec:descriptions-vague}

Successful projects rarely start with a well-defined problem.
Initial client descriptions are often vague, informal, or contradictory (P3, P6, P12), requiring practitioners to iteratively piece together objectives and constraints.
Elicitation thus involves ongoing engagement---meetings, document reviews, emails, and site visits with business and technical experts---to clarify assumptions and refine problem definitions (P3--6, P14--15).
P4 likened the process to a ``flywheel'' of iteration:

\begin{displayquote}
\textit{``You need to iterate [...] the whole process is like a flywheel that we go here, then probably we'll [...] formulate, come up with a solution, then put it out there, gather feedback and literally come back to [...] what's really your goal?''}
\end{displayquote}
Alignment on problem definition and success measures---for example, cost reduction, revenue increase, or efficiency---is crucial before translating the business problem into an optimization formulation (P10).
P10 emphasized asking, \textit{``how are we going to measure that we've delivered value?''}
P4 noted stakeholders respond best to specific, targeted questions, highlighting the need for focused engagement.
Despite its importance, this stage is time-intensive and often the longest, most challenging part of a project (P3--4, P6, P9).
P9 described a concrete instance of this flywheel on a university course-scheduling project: preference data from academic departments was not collected upfront, and its absence only became apparent once modeling was underway.
\textit{``Once we started solving the problem, then we kind of retroactively realized that we were kind of missing some of this preference information that would have been useful to formulate a nice objective that may have resulted in fewer of these change requests.''}
Recovering it meant looping back through non-technical stakeholders via \textit{``a pretty [...] low bandwidth medium of email and Excel,''} then manually reformatting it for the solver---\textit{``that's like a really slow feedback loop.''}
The gap surfaced only through modeling itself, not through any upfront elicitation checklist.

\subsubsection{Problem understanding requires reconciling distributed expertise}

Beyond sustained engagement, OMDs stressed the challenge of integrating knowledge fragmented across stakeholders.
Optimization projects involve business leaders, engineers, and end users, each with partial, sometimes conflicting perspectives (P4, P6, P9).
As P9 noted, \textit{``with this project in particular, it was the sheer number of stakeholders that we had to communicate with that I think was especially challenging.''}
Elicitation requires translating domain expertise into actionable constraints using non-technical language (P3, P6, P11).
Access to knowledgeable individuals can be a bottleneck (P15), and P10 emphasized \textit{``reading between the lines''} to uncover implicit needs.
Reconciling these perspectives is time-consuming but essential for a coherent problem definition and shared success criteria; without it, downstream modeling risks rework or rejection.

\subsubsection{OMDs critically assess when optimization is (and is not) appropriate}

OMDs stressed the need to determine whether optimization is the right tool or if simpler interventions---such as process changes or dashboards---are more appropriate.
As P10 noted, \textit{``sometimes the problem they may think is optimization, sometimes it's just about business process change.''}
This reflects a broader view of optimization as one option within a larger analytical toolbox, where understanding the problem precedes method selection.
Ensuring optimization will add real value requires a well-defined problem and clear expected impact. As P10 put it:

\begin{displayquote}
\textit{``It's because people are taking hammers and trying to find the nails. The answer is you have to find a nail and then pick out the right hammer.''}
\end{displayquote}
This played out concretely for P10 with a workforce-scheduling client who arrived believing they needed optimization.
A cost-benefit analysis instead traced the real problem to how task-duration data was being collected: workers were self-reporting time from memory at the end of the day rather than timing tasks directly, and \textit{``it turns out the majority of the times were divisible by five minutes.''}
P10's recommendation was not a model but a process fix: \textit{``Have a start and a stop button on your iPads, and then you'll record better times and save money. That's not analytics.''}
Determining that optimization was \textit{not} the answer was itself the analytical work---and it required the same investigative rigor as building a model would have.

\subsection{Data Processing}\label{sec:data_processing}

\begin{displayquote}
    \textit{``In terms of the total number of lines in my code base, I would say it was 70\% data engineering, 30\% of everything else. So data engineering is really the big pain point.'' \\ --- P1}
\end{displayquote}
After clarifying the problem, OMDs must assess the data landscape.
As P7 stated, \textit{``we need to understand the data that's going to be available before developing the model.''}
The data processing stage is often a major bottleneck, involving access, cleaning, and documentation of messy, sparse, or poorly described datasets (P3, P10, P13--14).
This confirms a well-established finding from data science practice, where data work is similarly iterative and costly~\citep{muller2019data}.
What goes beyond that prior literature is what the data is \textit{for}: in data science, data preparation feeds a learning task whose structure is already fixed, whereas here it determines the structure of the optimization problem itself (see also Section~\ref{sec:ml_comparison}).
Understanding the client’s data flows and technology stack is essential.
Early steps often include gathering sample data, running simple simulations, and producing documentation to ensure transparency and consistency (P10, P11).

\subsubsection{Data tends to be messy, sparse, or incomplete}

OMDs consistently described data quality as a major bottleneck.
Data is often messy and requires extensive cleaning and pre-processing, with some estimating that 70\% of their time goes to this step (P13, P14).
Data is also frequently sparse or incomplete, complicating robust modeling (P10).
For example, probabilities needed for stochastic programming are often unavailable, which forces simplified assumptions or reliance on domain experts (P3, P10).
Even after processing, concerns about accuracy and reliability remain.
As P7 reflected, \textit{``you're modeling a real life situation and you know, the data aren't always accurate [...] so it's, you know, how good do you have to be anyway?''}
In some cases, synthetic data offers a practical workaround when real datasets are incomplete or missing: \textit{``If [...] I could get simpler, cut-down versions that are not quite right but they look realistic enough [...] that person wouldn't have been a bottleneck''} (P6).
Though labor-intensive, these activities are essential to align data with optimization requirements and advance the project.

\subsubsection{Accessing and interpreting data is slow and difficult}

Challenges with messy, sparse, and incomplete data are intensified by unclear or missing documentation (P14), which requires substantial effort to interpret dataset meaning and flows.
Even producing simple sample data can be difficult due to access constraints (P3, P6, P9, P14).
Data access is often slow and cumbersome, which creates bottlenecks that delay testing and overall progress.
As P1 explained, \textit{``their data is not in a single place, or it's like, layered on top of a bunch of legacy systems. And so also separating that out was like a pretty big piece.''}
Because acquisition spans multiple teams and infrastructures, it requires extensive coordination and manipulation, with OMDs documenting structures, mapping data flows, and navigating technology stacks as essential groundwork for modeling (P10, P14).
Model development and data acquisition often proceed in parallel, creating feedback loops where data gaps stall modeling and modeling needs generate new data requirements (P3, P6).
Synthetic data may temporarily unblock development (P6), but OMDs emphasized that accessing and interpreting data remains one of the most persistent challenges in practice.

\subsubsection{Data drives problem definition and modeling}\label{sec:data-problem}

A consistent finding is that the problem cannot be fully formulated until the data is understood.
As P7 noted, \textit{``a lot of times the data will drive the knowledge, the understanding of the problem, frankly.''}
OMDs emphasized that optimization is shaped by the available data, often requiring a ``data-first'' rather than ``model-first'' approach (P10).
As P10 put it, describing a shift in their own practice: \textit{``Go from model first [...] to being data first. Worry about your data. The model comes second.''}
This is a stronger claim than ``data matters a lot in optimization too'': the data does not merely calibrate a pre-specified model, it determines which sets and indices exist, what is static versus dynamic, and which constraints are even expressible in the first place.

Early steps include distinguishing static from dynamic information, gathering representative samples, and running simplified simulations (P10, P11).
Mapping data flows, investing in pre-processing, and collecting samples not only enable model development but also help concretize the problem: they shape assumptions and sometimes prompt updates to the initial framing (P3, P5, P10).
Data collection and modeling typically proceed in iterative feedback loops: as data is acquired and analyzed, assumptions may be revised, techniques adapted, and even the problem formulation updated (P6).
Because data defines baselines, constraints, and feasible solutions, OMDs stressed understanding its role (P10).

\subsection{Model Development}

\begin{displayquote}
    \textit{``99\% of the time, something was not included in the model from the customer standpoint. They'll say, well, we have this, you can't do that [...] And so then you have to go back and update the model.'' \\ --- P7}
\end{displayquote}
Based on prior ML and data science workflow literature~\citep{wang2019human}, one might have expected optimization model development to resemble the comparatively standardized ``inner loop'' of algorithm selection and tuning described there.
However, model development in optimization is a distinct phase that differentiates it from most ML workflows (see also Section~\ref{sec:ml_comparison}).
Unlike ML---where prediction tasks can often be operationalized without formal mathematical specification---OMDs must translate natural-language descriptions into mathematical models (P10, P12, P14--15).
P10 drew this contrast to ML directly: \textit{``If you take the machine learning types of things, there's documentation out there that basically says if you take this data and you throw it in here, you're going to get this answer [...] you don't worry about what's under the hood.''}
Optimization, they argued, instead requires first discovering what the decision, its constraints, and its measure of quality even are.
This upstream elicitation work has no direct ML analogue, since ML's objective (i.e., a loss function or target metric) is typically fixed by the task framing rather than negotiated stakeholder by stakeholder.

This translation is demanding: practitioners must balance tractability, realism, and theoretical soundness while working with both data and stakeholders (P10, P13--14).
The challenge is amplified by the fact that many valid ways to model the same problem exist, with different formulations yielding dramatically different computational properties~\citep{IPref}.
As P11 noted, \textit{``modeling is quite an art.''}
OMDs therefore rely on prior models and the literature, adapting and simplifying them to context while making assumptions explicit and validating them with stakeholders (P4--5, P8, P10--11, P14).
Early work often involves minimum viable models that are iteratively refined through testing, feedback, and deeper data engagement (P10, P12--14).
This resource-intensive process is a challenge unique to optimization.

\subsubsection{Modeling is iterative and evolves with shifting requirements}\label{sec:translate-math}

Translating vague or high-level business requests into precise mathematical formulations is one of the most challenging and time-consuming parts of the workflow (P12, P14--15).
As P12 noted, \textit{``it's a lot of figuring out what he wants [...] and then we have to somehow translate that into math.''}
Because requirements evolve, modeling is inherently iterative (P4, P13--14), with algorithmic choices adapting as understanding deepens (P4, P7, P13).
Navigating this evolving process is essential to producing models that are technically sound and aligned with organizational needs (P14--15).
To manage uncertainty, practitioners adopt agile strategies \citep{sharp2006agile,sy2008optimizing}, starting with minimum viable models and refining them through successive iterations (P10, P14).
Stakeholder feedback and data realities guide each revision (P4--5, P10).
As P5 explained, stakeholder requests are typically underspecified: \textit{``It’s never specific enough from a modeling perspective.''}
This requires documenting assumptions and continuously verifying alignment with stakeholder expectations and implementation needs (P10).
The effort is substantial---sometimes exceeding budgets or extending over months of iteration (P7, P10, P12, P15).

\subsubsection{OMDs balance simplification with realism}\label{sec:simplification-realism}

Modeling requires creativity and abstraction to simplify real-world complexity without losing essential features.
OMDs must balance fidelity with solver runtime, stakeholder priorities, and data constraints, blending simplification and realism to keep models both sound and useful.
Rather than pursuing strict optimality, they often \textit{``satisfice,''} aiming for solutions that are good enough within practical runtime (P4--5).
As P13 noted, \textit{``if you could push something out in a day, it’s better than a week,''} highlighting the speed-perfection trade-off.
This approach involves abstracting problems, deciding which constraints can be simplified, and iteratively testing on available data (P3--4, P15).
Critically, this balancing act includes checking the formulation itself against reality while it is still being built, not only validating a solution once one exists (cf.~Section~\ref{sec:validation}). 
P3 described building fidelity up in stages, checking in with stakeholders as the formulation grew more detailed: \textit{``We don't need to really model it exactly to get their opinions. We just need to know what are the things we want to include [...] for some other parts we need to really think more about it [...] we model it more exactly and we ask them whether it makes sense or not [...] you need to check the assumptions with them to see whether they make sense for their context of the problem.''}

These checks may sometimes surface a formulation error directly, before any solution is ever computed: P10 described discovering that a stakeholder-reported ``hard'' constraint was in practice routinely violated, which required the constraint itself to be revised from hard to soft: \textit{``They may have told you we always do this, but then you find in the real world that five times they don't [...] you have to start building that into your model.''}
The tractability-realism trade-off itself sometimes had no single right answer.
P13 described formulating a spatial-separation constraint two different ways---a theoretically complete encoding of the full space, and a simpler ``engineering'' proxy---with neither fully resolving the issue on its own: \textit{``We would just find ourselves coming right back to the same issue, or realizing that we had over-complicated the same technique.''}
Rather than picking one unilaterally, P13 presented both options to the stakeholder alongside numerical experiments showing where each broke down.
The trade-off was resolved as a negotiated decision, not a purely technical one.

\subsubsection{OMDs leverage literature and expert judgment}\label{sec:lit-expert-judgment}

OMDs frequently draw on both the literature and expert judgment when translating vague or high-level requests into formal models (P12, P14--15).
They typically start by identifying related model types from prior work (e.g., knapsack, scheduling) and adapt these structures to the specific context---a process often described as ``\textit{scaffolding}'' (P4--5, P8, P14).
Literature and graduate coursework (P12) provide reusable components such as variable definitions, common constraints, and solution strategies, which help accelerate development and ground models in established practice (P4--5, P11).
Initial modeling also relies on practitioner experience to fill theoretical gaps (P12, P14--15).
As P8 explained, \textit{``I read a few different papers related to VRPs with pickup and drop-off windows and [...] my model was like an amalgamation,''} illustrating how OMDs synthesize insights from multiple sources to build tailored solutions.
Scaffolding further enables use of specialized solvers that outperform general-purpose alternatives~\citep{parmentier2022learning}.

\subsection{Model Implementation}\label{sec:model_implementation}

\begin{displayquote}
    \textit{``It was a mess. We had to open the documentation and try to understand how it works and how you could do the callback and what you pass to the callback and everything.'' \\ --- P4}
\end{displayquote}
Once an initial model is formulated, OMDs move to implementation (see Figure~\ref{fig:artifacts} in Section~\ref{sec:problem-elicitation}).
This stage is often time-consuming and requires translating formulations into specialized code and extensive data handling.
Experiences vary: some view it as a straightforward technical task that \textit{``doesn’t use a lot of brain power''} (P4), while others see it as a persistent bottleneck.
As P8 noted, \textit{``I feel like I spent a lot of time getting stuck in things saving properly or loading to the wrong environment.''}
We expected this stage to be comparatively mechanical, following software-engineering norms for translating a specification into code.
That expectation was only partly confirmed: iterative prototyping does resemble general software practice, but implementation is also where OMDs confront solver-specific opacity that has no clean software-engineering analogue---debugging against a black-box solver, not just against their own code (see also Section~\ref{sec:ml_comparison}).
Despite different experiences, OMDs agreed that implementation is essential, both to operationalize the model and to validate assumptions about data quality and scalability.

\subsubsection{Implementation advances through iterative prototyping}

OMDs described implementation as incremental and pragmatic.
They often start by prototyping on toy problems, testing \textit{``constraint by constraint''} before scaling up (P5, P8).
Early implementations are treated as minimum viable products, prioritizing correctness over speed or efficiency (P4, P12).
Many draw on existing resources, such as pseudocode or open-source implementations, and adapt them to their solver environments.
As P5 noted, \textit{``sometimes pseudocode is the best method [...] starting from open source or papers with code gets you the first leg up.''}
This scaffolding mirrors model development, in that practitioners map new problems onto known formulations rather than starting from scratch.

\subsubsection{Initial implementations validate the scalability of the proposed model} \label{sec:validate-scalability}

Once an initial implementation correctly encoded the constraints, OMDs used it to test whether a formulation could scale.
A common strategy was running the model on off-the-shelf solvers, which quickly revealed tractability issues.
P13 and P15 reported models unsolvable on local machines, while P10 recalled one that took 48 hours, making iteration difficult.
P12 described a similar challenge: \textit{``We started out with a single model and the runtime would be over 24 hours or so, which we definitely could not do.''}
To address bottlenecks, developers used various tactics.
Some improved solution quality with heuristics, others reformulated models to reduce computational costs (P4, P14).
P12 broke the problem into stages and added heuristics to achieve acceptable performance.
P15 used a \textit{relaxation}, an ``easier'' version with fewer constraints, to generate feasible solutions when the full formulation was too costly.
P14 reformulated the model more fundamentally in response to computation: an arc-based MIP formulation for a large-scale vehicle-routing problem worked at small scale but was \textit{``impossible to solve by commercial solvers''} once the problem grew to production size, prompting a switch to a path-based, column-generation formulation instead.
These adjustments show how scalability testing guided decisions on heuristics, decomposition, and more efficient formulations.

\subsubsection{OMDs struggle to leverage off-the-shelf optimization solvers}\label{sec:off-the-shelf}

This is a distinct frustration from the scalability challenges above: even when a solver could handle the problem size, OMDs found the solver itself difficult to work with as a piece of software.
Despite their widespread use, OMDs reported frustrations with off-the-shelf solvers.
Several OMDs (P2, P11--12, P15) struggled with limited documentation and complex solver websites.
Modern MILP solvers, including Gurobi~\citep{achterberg2019s} and CPLEX~\citep{nickel2022decision}, come with a staggering number of parameters that control key algorithmic components of the solver, such as cutting plane separators, branching rules, and heuristics.
These parameters can greatly affect runtime but are notoriously hard to tune (P2, P11--12, P15).
As P12 explained:

\begin{displayquote}
    \textit{``[Gurobi] have their own internal parameters that we tried out, and you could set a time limit and things. So, we tried to understand how Gurobi functioned a little bit, but it's still like a black box.''}
\end{displayquote}
This frustration went beyond parameter opacity alone and recurred in (at least) two specific forms.
First, parameter semantics are often unexplained, forcing guess-and-check rather than principled tuning.
P15 elaborated: \textit{``It's very difficult to get a high level conceptual idea of how different parameters might affect your model and solving your model. So really the only thing to do is kind of just tweak the parameters and see what happens [...] a lot of guess and check.''}
Second, documentation itself was unstable, not just sparse: P11 described documentation links moving without notice---\textit{``you go there and then it's just like missing''}---forcing them to reverse-engineer usage from code examples instead.
Frustration increased when using advanced features such as callbacks (P4, P15). 

Navigating the correct parameters is a much more nuanced task than analogous processes in ML workflows \cite{shankar2024we}.
P5 compared multi-objective optimization to ML hyperparameter search as follows: \textit{``Usual hyperparameter settings---you have some output you care about, like machine learning, you do hyperparameter selection to maximize out-of-sample accuracy [...] when I'm thinking about multi-objective optimization, it's a little different because all of those solutions are kind of equally valid if you have Pareto-optimal solutions.''}

\subsection{Model Validation} \label{sec:validation}

\begin{displayquote}
    \textit{``Do you think it makes sense? Do you think this kind of managerial insights are really working for this problem? Is it what you observe during your work?'' \\ --- P3}
\end{displayquote}
After implementation, the next stage is validation---ensuring the system is correct and useful.
This involves three questions: $(i)$ does the model capture the real-world decision problem; $(ii)$ does the code correctly implement it; and $(iii)$ can the approach scale to required instance sizes.
Addressing the first requires collaboration with domain experts, while the latter two are typically handled by practitioners via manual checks, ad-hoc tests, and performance experiments.
Prior literature on MLOps would lead us to expect validation centered on quantitative, automated metrics computed against held-out data~\citep{shankar2024we}.
We found the opposite emphasis: validation here is interpretive and collaborative because optimization data has no equivalent of a ground-truth label to automatically score against, so manual sniff tests and stakeholder critique carry the weight that an automated evaluation pipeline would in ML (see also Section~\ref{sec:ml_comparison}).

\subsubsection{Models are validated through inspection and stakeholder critique}\label{sec:sniff-tests}

Validation in practice was highly interpretive and collaborative, involving both developers and domain experts.
Manual sniff tests ranged from checking small instances (P14) to visually spotting errors, such as infeasible pickup and drop-off times (P8).
Some supplemented these checks with auxiliary metrics or sensitivity analysis (P4, P11), while a few used formal processes like independent \textit{``solution validators''} (P10).
The rigor of these checks varied widely across practitioners.
At one end, P9 built automated assertions: \textit{``Unit tests that were asserting some kind of standard things that should be true [...] sanity checks that this schedule was only scheduled once, that this exam was only scheduled once, that all the exams were scheduled.''}
At the other end, P13 kept no formal test suite at all, relying instead on memory: \textit{``I do not care about edge cases at all [...] it's not a test suite that's automated. It's just---here's a couple problems that I know in my head I should keep running [...] if I was to come back to some of the work I've done, I wouldn't know how to actually test and validate it anymore.''}
For decomposed problems, checks often happened at the sub-problem level before scaling up: P8 described \textit{``very rudimentary checks for those sub-problems''} (e.g., time-window feasibility, visual cycle-checking), followed by the harder question of whether a locally valid piece would \textit{``actually help when I apply these edges to the bigger problem.''}
Domain experts reviewed candidate solutions, surfacing missing requirements or mismatches.
As P9 explained, \textit{``showing somebody a solution and saying, `okay, this is what we think we’re going to do,' is kind of the best way for them to look at it with a critical eye.''}
Feedback was iterative, with teams running scenarios and refining models (P7, P9).
For larger instances, which could feel like a \textit{``black box''}~(P13), developers relied on small examples and domain-specific visualizations---from CAD drawings for engineers (P15) to sample course schedules for administrators (P9).

Developers also produced \textit{``lots of visualizations''} (P10) to aid debugging.
This translation work went beyond generic visualization.
P1, asked directly how they convinced a client the model was right, described reformatting the output to match an artifact the client already trusted: \textit{``The biggest thing was they had certain sheets or summaries of the system that they were used to seeing every day, and putting it in that format [...] basically showing them a mockup of what it would be in a format they are very used to is something that did a lot of work.''}
P13 took a complementary approach, exposing the mechanism rather than just the answer: a visualization plotting the solver's optimality gap against time, with the contribution of each competing objective color-coded; they also explicitly \textit{``had to explain the cause and effect of every type of input.''}

A related incident shows this same translation work shaping how solutions were consumed, not just explained: P13 described a client who rejected an early version of a decision-support tool because it withheld this kind of legibility, calling it a \textit{``magical box''} that took hours to return an answer with no way to intervene.
In response, P13 showed the solver's optimality gap and time limit and built a callback that let stakeholders inspect an early, partial solution instead of waiting for the final one.
This made the specific solution transparent and inspectable rather than relying on an overall performance metric, which helped build trust.
These examples underscore that this kind of translation is only possible because the underlying model is explicit in the first place.
Inspection, critique, and visualization together were the primary mechanisms for verifying model behavior and producing stakeholder-trusted solutions.

\subsubsection{OMDs satisfice with respect to computation times}

Unlike academic literature, OMDs treated runtime as a practical constraint rather than a quantity to minimize.
They adopted a \textit{satisficing} approach: solutions delivered within a stakeholder-defined budget---for example, fifteen minutes for monthly planning---were acceptable (P14--15).
As P15 noted, solutions only needed to be \textit{``fast enough to meet our business needs.''}
Once this threshold was met, developers focused on balancing accuracy and runtime for \textit{``good enough''} performance (P4, P8).
Validation of computation times thus prioritized meeting practical stakeholder needs over theoretical efficiency.
These budgets were typically derived from an operational cadence rather than chosen abstractly. P14 explained: \textit{``We do have some requests from them about how long this algorithm should take. Because this needs to be run every day [...] we want to have an algorithm that can solve in 15 minutes. That's our target.''}
Once set, these expectations were highly contextual.
As P4 put it: \textit{``It depends on the usage [...] if this is a planning MIP that is run every once a month and the customer is fine waiting 15 minutes, I'm okay with solving it in 15 minutes.''}
These expectations are not always given upfront, however.
P13 described a project where the client initially expected a model solved to optimality but, through repeated iteration, converged with the practitioner on a much looser bar---exposing a callback that stopped once the most important constraints were satisfied rather than waiting hours for full optimality: \textit{``That was a lot of back and forth trying to figure out what exactly they wanted, because they also struggled to know what could be done to them.''}

\subsubsection{Model validation is concurrent with data validation}

OMDs stressed that validating models is inseparable from validating the data driving them.
Much debugging focused on aligning model assumptions with real-world data.
P7 noted that, lacking suitable test cases, developers sometimes had to \textit{``write a problem generator''} to explore scenarios.
Practical barriers to accessing representative data were common; P15 described difficulties obtaining test data from clients.
Several OMDs built mechanisms to check whether data assumptions held before trusting outputs.
For example, P10 implemented a \textit{``data validator''} that flagged violations, noting the process truly began once real data was used.
When such mismatches occurred, validation often required revisiting both pre-processing and the model: \textit{``Iterate back if assumptions are not met and potentially modify the data prep or modeling''} (P10).
Two edge cases P10 described illustrate what this validator actually caught in production.
First, a schema assumption failed silently: an internal record ID was assumed to always exist, so records missing it were being dropped without anyone noticing---\textit{``sometimes that internal ID hasn't been created [...] you don't give us the internal ID, we're dropping that record and we're just going on''}---until a stakeholder caught the loss and the team began synthesizing placeholder IDs instead.
Second, unexpected input broke the implementation outright: stakeholders could specify a scheduling preference for a shift that did not exist in the system, \textit{``they broke our code. Right. We never saw that before.''}
Both were added to the data validator only after production use surfaced them, not anticipated in advance.

This validation practice also highlights a contrast with data validation in ML: ML/MLOps data validation checks whether new data still conforms to existing expectations (e.g., schema, bounds, completeness) that an already-fixed model was built to expect, which is generic enough to automate once and run on a schedule~\citep{shankar2024we}.
Optimization data validation instead checks whether data, together with the model's constraints, still yields a feasible and sensible decision; because data determines the structure of the problem itself (see Section~\ref{sec:data-problem}), there is no comparably reusable check, since what counts as valid data is redefined by each problem formulation.

\subsection{Deployment}

\begin{displayquote}
    \textit{``When I talk about deployment, it's not just standing up an application technically. It's also about [...] what change management do you have to do to be successful in the business? How are people going to use this application? How are they going to change the way they're working?'' \\ --- P10}
\end{displayquote}
MLOps literature would lead us to expect that deployment centered on continuous integration, automated monitoring, and retraining cadences~\citep{shankar2024we,kreuzberger2023machine}.
We found a different philosophy altogether, organized around documentation, change management, and portability rather than automated pipelines (see also Section~\ref{sec:ml_comparison}).
``Deployment'' covered three distinct senses in our interviews, differing in what handoff actually meant in practice---OMDs described variation in the final stage of projects, depending on use case and context.
First, deployment as a static artifact.
For example, a schedule, routing plan, or allocation that stakeholders could act on (P9, P12--13).
As P12 noted, \textit{``the end product is what they looked like \emph{[sic]} and were able to understand,''} highlighting that success meant outputs were interpretable and usable.
Second, deployment as a proof of concept handed off to another team.
P13 and P14 described projects where demonstrating feasibility was enough, while integration and scaling were done by production engineers.
Third, deployment as full integration by the OMDs themselves.
Beyond technical handoff, successful deployment requires organizational coordination.
P10 emphasized aligning stakeholders on adoption, including user training and support.
Lifecycle management---monitoring, assessing value, and updating systems---was sometimes performed by implementers and sometimes transferred to clients or handled jointly.

This coordination could expand mid-project in ways the original problem framing never anticipated.
P10 described a scheduling solution that worked well enough to reduce headcount needs, pulling an entirely new stakeholder into a project that was already underway: \textit{``If we end up with a scheduling solution that is going to be scheduling people to do work and it turns out that they need less people to do the same amount of work [...] that means you're going to be doing layoffs. We have to have a conversation with the HR department. They weren't a stakeholder at the beginning. We didn't anticipate the fact that this would be so good.''}
A model's own success, not a technical failure, was what forced the scope of deployment to widen.

\subsubsection{Documentation is key to successful deployments}\label{sec:doc-key}

OMDs emphasized that documentation and communication are essential for adoption but time-consuming.
As P4 noted, \textit{``I would say that writing the document itself takes [...] time.''}
Deployment requires materials tailored to diverse stakeholders, often including problem-specific interfaces with metrics or visualizations to make results actionable (P4, P9).
Documentation supports training and builds trust, especially since optimization is often seen as a \textit{``black box''} (or, as P13 described it, a \textit{``magical box''})---here, in the stakeholder-facing sense discussed further in Section~\ref{sec:ml_comparison}, not OMDs' own relationship to solver internals.
Visualizations are crucial for conveying outputs and demonstrating alignment with stakeholder expectations, which ultimately fosters transparency and confidence (P13--14).

\subsubsection{Portability and integration constraints shape deployment}\label{sec:portability}

Unlike ML, where core tools and libraries are largely open-source, optimization depends heavily on commercial, proprietary solvers.
This creates unique barriers around transparency, reproducibility, and access that shape how practitioners work.
OMDs described frequent challenges deploying models across diverse technical environments, as client restrictions or system limitations often dictated available tools.
In some cases, commercial solvers were prohibited, which required teams to approximate solutions with less capable tools.
P3 recalled: \textit{``They are not trained as experts for operations research. So they don't quite use Python, Gurobi, or any of it [...] so we decided to build a Microsoft Excel tool for them.''}
Portability issues also affected solvers (P15) and integration with third-party platforms, where missing features or incompatible interfaces created barriers (P1, P11).
These examples show that deployment is often shaped more by technical and organizational constraints than the model itself.
This ultimately forces OMDs to adapt solutions to client-approved environments, even at the cost of functionality.

Solver choice was frequently an explicit, upfront negotiation instead of a purely technical decision.
P7 described clients who, anticipating a future commercial product, ruled out Gurobi~\citep{achterberg2019s} outright: \textit{``We don't want to buy a Gurobi license, so don't use it.''}
P14 described a technology-transfer project that required avoiding commercial solvers entirely in favor of open-source alternatives, since the receiving organization \textit{``might not have the commercial license.''}
P5 described the organizational cost of licensing itself, independent of a tool's quality: \textit{``Nobody wants to go through all the individual approvals of any new software onto a network [...] it's easier for them to trust you and you build something yourself than it is for them to go spend an unspecified amount of time approving someone else.''}
P10 described a financial-services client whose environment restrictions required maintaining a translation layer between Python (the OMD's preferred tooling) and Java (the client's approved stack).
These are recurring, concrete deployment costs with no direct counterpart in ML's more open, portable tooling ecosystem (e.g., scikit-learn~\cite{pedregosa2011scikit} or PyTorch~\cite{paszke2019pytorch}).
\section{Discussion} \label{sec:discussion}

Our findings show that optimization model development unfolds as a highly iterative and socio-technical process.
Across the workflow, three themes consistently surfaced as central to optimization practice: data, decisions, and dialogue.
In the discussion that follows, we unpack each of these themes and highlight opportunities for future tooling that can better support OMDs.
We also discuss how optimization differs from previously studied ML workflows.

\subsection{Data, Decisions, and Dialogue: Three Ds of Optimization Practice}

Findings from our study suggest three broad themes that characterize successful optimization workflows: data, decisions, and dialogue.
Please refer back to Figure~\ref{fig:workflow} in Section~\ref{sec:findings} for a visualization of the whole workflow.

\subsubsection{Data}  

In ML, data preparation is a distinct pipeline stage that feeds a learning task whose structure is already fixed---the target variable, model class, and evaluation metric do not change based on what the data turns out to look like \cite{shankar2024we}.
In optimization, by contrast, data does more than parameterize an already-specified model: it determines the mathematical problem's structure itself---which sets and indices exist, what is treated as static versus dynamic, and which constraints are even expressible.
Our findings present data processing as its own stage, but one that remains tightly coupled to model formulation throughout the workflow.
As our findings show, data wrangling alone can consume the majority of project effort and shapes both formulation and validation.
And yet, optimization education still treats data as clean, provided, and secondary.
This mismatch between how optimization is taught and how it is practiced is itself a design and pedagogical opportunity.

This gap resonates with \citet{sambasivan2021everyone}'s account of data cascades in high-stakes ML: underinvestment in data work rarely produces immediate, visible failures, but instead compounds silently until it surfaces downstream; often in ways that are costly to diagnose and repair.
Our findings suggest an optimization-specific variant of this risk: because data does not merely feed a model but actively shapes problem formulation (Section~\ref{sec:data-problem}), a data cascade in optimization does not only degrade output quality (as in prediction tasks) but can propagate into the wrong decisions being formalized as constraints or objectives.
Left unaddressed, an early data gap can therefore resurface much later---during validation or even deployment---as a mismatch between the model and the real-world decision it was meant to support.
This ultimately requires OMDs to backtrack across multiple workflow stages instead of simply retraining or patching a model.

Our findings echo observations from ML that data preparation is iterative and costly \cite{muller2019data}, but optimization introduces unique challenges.
First, data shapes the form and scope of the problem itself---for example, whether to optimize over a daily, weekly, or monthly horizon---rather than directly driving model performance.
Second, optimization often relies on unlabeled data for one-off or periodic decisions, in contrast to the labeled datasets and continuous train/test loops of MLOps~\citep{shankar2024we}.
By ``label'' we mean a ground-truth target value---the correct classification or outcome a supervised model is trained and automatically evaluated against.
Optimization input data (e.g., demand figures, travel times, cost parameters) has no equivalent: there is no held-out annotation a practitioner can compare against to score data quality automatically.
A consequence follows directly from our findings: because there is no scalable, automatic correctness signal, data errors surface only through manual, iterative checks against the model and stakeholder judgment (e.g., the sniff-test and ``iterate back if assumptions are not met'' practices we document) rather than through an automated evaluation pipeline.
Such errors can also go undetected for longer, since nothing forces the check to happen on a fixed cadence the way a train/test loop in ML does.
This makes tooling designed for ML pipelines---focused on labeling \cite{ratner2017snorkel}, monitoring drift \cite{mallick2022matchmaker}, or retraining \cite{diethe2019continual}---poor fits for optimization.
Finally, an important exception arises when data encodes uncertainty, such as stochastic travel times or fluctuating demands.
In these cases, richer or more accurate data can materially improve decision quality, motivating lines of research on integrating predictions into optimization models (see \cite{mandi2024decision} for a discussion).
The central role of data underscores the need for tools that treat it as a first-class concern, not just an input to solvers (see potential directions in Section~\ref{sec:tooling}).

\subsubsection{Decisions}

Our interviews reinforce that the core goal of optimization is to support better decisions.
While research has long focused on algorithms for hard decision-making problems \cite{wolsey2020integer}, practitioners emphasized that models are only valuable if they yield decisions that are timely, interpretable, and actionable.
This pragmatism pervaded the workflow, as OMDs balanced model fidelity, computational efficiency, and data availability in ways that reflected organizational realities.

A central strategy was \textit{satisficing}---choosing solutions that are ``good enough'' rather than strictly optimal.
Computation time and model fidelity were treated as levers to be traded off: problems might be simplified into well-known classes with specialized solvers \cite{parmentier2022learning}, or heuristics substituted for formal methods when speed was crucial (Section~\ref{sec:findings}).
Tools that foreground such trade-offs and let users explore multiple modeling paths---from high-fidelity but slow to lightweight and fast---are better aligned with the decision-making realities practitioners face (more in Section~\ref{sec:tooling}).

Automation (with its promise and perils \cite{wang2019human}) is a natural extension of these trade-offs: it could relieve bottlenecks in data cleaning, generate solver-ready formulations and code, or check the feasibility of candidate solutions.
Our interview protocol, however, was not designed to systematically probe current AI or LLM tool adoption among OMDs, which remains an important direction for future work.
The promise and perils of automation connect to recent research showing that users often over-rely on automated outputs, even when they are flawed \cite{schoeffer2024explanations,swaroop2025personalising,qiao2025use,buccinca2021trust}.
Some design choices---such as surfacing assumptions, exposing inconsistencies, or citing sources---have been shown to mitigate this over-reliance~\citep{kim2025fostering,raees2026people}.

Together, these findings point to a crucial design imperative: optimization tools should augment rather than automate away human expertise.
By making trade-offs visible, supporting satisficing, and embedding features that encourage critical evaluation, decision-support tools can scaffold better decisions---not just faster ones.

\subsubsection{Dialogue}\label{sec:dialogue}

Dialogue is a foundational mechanism throughout the workflow, integrating input from engineers, business experts, and end users into problem definitions, constraints, and objectives.
This iterative, often non-linear process mirrors the ``outer loop'' described by \citet{kross2021orienting}, in which data scientists engage continuously with clients; similarly, OMDs interact with stakeholders throughout problem framing, model development, and validation to ensure that technical solutions align with organizational goals.

Yet, dialogue is frequently slow and time-intensive---a major bottleneck, since, as \citet{delarue2024algorithmic} note, exchanging information across expertise levels further slows progress.
Novel tooling approaches should therefore prioritize accelerating dialogue without sacrificing its accuracy.
Further details are provided in Section~\ref{sec:tooling}.

As our findings show, visualizations, concrete examples, and simplified scenarios extend this dialogue, letting stakeholders reason about intermediate results without deep technical expertise.
Transparency is closely intertwined with this dialogue: by making assumptions, intermediate results, and potential limitations visible, OMDs help stakeholders understand the model’s reasoning and build trust~\citep{ferrario2022explainability,shin2021effects,lee2004trust}.
``Sniff tests'' (Section~\ref{sec:sniff-tests}) help facilitate transparent communication and reduce the perception of optimization as a ``black box''---in the stakeholder-facing sense unpacked further in Section~\ref{sec:ml_comparison}---and align with broader challenges in AI transparency~\citep{liao2024ai}, including the need to tailor explanations to individual stakeholder backgrounds and needs~\citep{ehsan2021operationalizing,ehsan2024xai}.
Proper interface design is crucial in this context to prevent unwarranted trust while enabling stakeholders to engage meaningfully with the model and its outputs~\citep{schlicker2025we,lee2025veriplan,leiser2024hill}.
Ultimately, well-designed dialogue and interfaces are key to fostering both understanding and appropriate reliance~\citep{schemmer2023appropriate} on complex optimization systems.

\subsection{Optimization vs. ML Workflows} \label{sec:ml_comparison}

We use ML workflows not as a benchmark against which optimization should be judged, but as an analytical contrast that helps identify which aspects of optimization practice require distinct forms of support.
Prior research on ML workflows has established a vocabulary for understanding challenges such as data preparation, validation, documentation, and deployment.
Our findings show that these activities take on different meanings in optimization.
The optimization workflow is a highly iterative and socio-technical process.
While this progression shares the iterative, multi-stage structure seen in other data-intensive workflows---as in CRISP-DM~\citep{wirth2000crisp} and MLOps frameworks~\citep{shankar2024we}---the optimization workflow is fundamentally differentiated along several dimensions we unpack below: the upstream labor of discovering an objective and its constraints, the structural (not merely parametric) role of data, the fact that outputs are typically one-off and human-inspected before anyone acts on them, the demands an explicit and auditable model places on stakeholder dialogue, and distinctive deployment and portability constraints.

A first key distinction lies in the desired outcome of the work.
Optimization is designed to generate actionable decisions by selecting the best option from alternatives under situational constraints, providing prescriptive results.
Conversely, the majority of ML and data science pipelines focus on prediction, aiming to extract knowledge and insights from data~\citep{wang2019human} to support descriptive or predictive outcomes~\citep{shankar2024we}.
MLOps workflows, such as those studied by \citet{shankar2024we} and \citet{kreuzberger2023machine}, are aimed at continuously generating predictions in an automated fashion and maintaining production ML systems reliably and efficiently.

This divergence in goals leads to a strikingly different technical core in the modeling phase.
In optimization, model development requires the specialized expertise of translating high-level, ambiguous business needs into an explicit and typically highly customized mathematical optimization model (see also Figure~\ref{fig:artifacts} in Section~\ref{sec:problem-elicitation}), such as an MILP.
This translation relies heavily on academic literature, expert judgment, and prior models as ``scaffolding'' as different modeling choices can often have dramatic impacts on the computational tractability of the model.
As our findings describe (Section~\ref{sec:findings}), OMDs must first discover what the decision, its constraints, and its measure of quality even are before any modeling can begin.
This upstream elicitation work has no direct ML analogue, since ML's objective (e.g., minimizing a loss function) is typically fixed by the task framing rather than negotiated stakeholder by stakeholder.
This specialized mathematical modeling is a unique, resource-intensive effort within the OMD process.

It also presents distinct challenges during model implementation, where writing software to solve an optimization problem typically requires implementing custom logic to encode the model and iterative prototyping to ensure it is implemented correctly. 
In contrast, the technical inner loop (``magic''~\citep{kross2021orienting}) of data science collaboration or the \textit{Modeling} phase of the technical workflow from \citet{wang2019human} focuses on statistically driven tasks, such as algorithm selection, hyperparameter optimization, and ensembling.
These tasks generally adhere to standard protocols, streamlined implementations, and are more suited to automation~\citep{wang2019human}.
Moreover, model validation in ML tends to rely more heavily on quantitative metrics, such as predictive accuracy for supervised models~\citep{shankar2024we}, whereas model validation in optimization focuses on ensuring that the model adequately captures the real-world problem and can be solved sufficiently quickly for practical use.

The role of data in optimization is substantially unique as well.
While data preparation is often acknowledged as time-consuming across all fields~\citep{muller2019data}, in optimization, data is a pervasive constraint that directly shapes the form and scope of the optimization problem itself.
The nature of the available data often dictates the knowledge and understanding of the problem.
As our findings describe (Section~\ref{sec:findings}), OMDs often shift from a ``model-first'' to a ``data-first'' approach.
This is a stronger claim than ``data matters a lot in optimization too'': the data does not merely calibrate a pre-specified model, it determines which sets and indices exist, what is static versus dynamic, and which constraints are even expressible in the first place.
As such, data processing and model development are tightly connected and interdependent in optimization.
In ML workflows, data scientists actively shape data through five types of interventions: discovery, capture, curation, design, and creation, to prepare it for analysis~\citep{muller2019data}.
Furthermore, in MLOps, data preparation is often automated and runs on a schedule to support continuous training~\citep{shankar2024we}.

A further distinction concerns how outputs are consumed.
An optimization output (e.g., a schedule, a routing plan, or an allocation) can be produced once or periodically and reviewed by a person before anyone acts on it; a production ML system, by contrast, typically generates predictions continuously and at a scale where individual outputs are not inspected~\citep{kreuzberger2023machine,xin2021whither}.
Our findings illustrate this concretely: P13 described a client rejecting an early, opaque version of a decision-support tool, which led P13 to expose the solver's optimality gap and build a callback so stakeholders could inspect an early, partial solution rather than wait for a final one (see Section~\ref{sec:findings}).
This pattern recurs throughout model validation: OMDs rely on manual sniff tests, domain-specific visualizations, and small illustrative examples precisely because a specific, inspectable solution is what stakeholders need in order to trust it.
MLOps tooling, by contrast, is built around continuous monitoring of aggregate behavior (e.g., service-level objectives, on-call rotations, or drift detection) because individual predictions are typically not inspected at production scale~\citep{shankar2024we}.

This connects to a second, related distinction: what an explicit, auditable model makes available to stakeholders.
Our participants used ``black box'' in (at least) three distinct senses that are worth separating.
The first is the solver's \textit{internals}: OMDs generally do not know exactly how commercial solvers search the solution space, much as an ML practitioner need not know how a package like scikit-learn~\citep{pedregosa2011scikit} works internally to use it productively (recall P10).
The second is a large problem \textit{instance} being hard for the OMD to grasp as a whole, independent of any solver or any stakeholder (recall P13).
The third, most consequential sense is the \textit{whole application} as experienced by a stakeholder or client (recall P13).
It is this third sense that optimization is distinctively positioned to address.
As \citet{le2025making} argue, optimization models are ``inherently interpretable in theory'' yet ``in practice, the scale, complexity, and constraints of real-world problems make it difficult to communicate the rationale behind optimization outcomes to stakeholders.''
Our findings show OMDs actively closing that gap: reformatting model output to match an artifact stakeholders already trust, proactively exposing a solution's cause-and-effect, and building independent ``solution validators'' that check outputs against a plain-language specification.
All these approaches convert a model that is transparent to the OMD into an artifact stakeholders can contest.
This is a different design opportunity than ML's post-hoc explainability agenda~\citep{dwivedi2023explainable}, which compensates for a model that is opaque by construction.
As a consequence, OMDs cannot fall back on treating their own model as a black box the way an ML practitioner can treat a trained predictor; they must develop and maintain a deeper working understanding of the problem domain itself.
We see this domain-understanding demand as a precondition for the heavy elicitation and dialogue documented throughout our findings.

Deployment philosophies also reflect the differing goals between ML and optimization.
MLOps workflows~\citep{shankar2024we,kreuzberger2023machine} emphasize automation through CI/CD, continuous monitoring of model performance, and automatic retraining cadences to maintain model validity over time (``repeated-use ML''~\citep{shankar2024we}).
Conversely, the success of optimization deployment depends heavily on extensive documentation and change management to explain the resulting outcomes (e.g., schedules or allocations) to stakeholders and manage portability constraints.
In this context, OMDs often favor a ``good enough'' solution (\textit{satisficing}) that meets the defined runtime constraints instead of pursuing strict optimality.
Portability constraints are distinctive in kind.
Production optimization commonly depends on licensed commercial solvers such as Gurobi~\citep{achterberg2019s} and CPLEX~\citep{nickel2022decision}, in contrast to the open-source tooling---scikit-learn~\cite{pedregosa2011scikit} or PyTorch~\cite{paszke2019pytorch}---that dominates ML deployment.
As our findings describe (Section~\ref{sec:portability}), solver choice was frequently an explicit, upfront negotiation, driven by anticipated commercialization, technology-transfer restrictions, and the organizational cost of licensing approval itself.

\subsection{Opportunities for Tooling and Support}\label{sec:tooling}

The opportunities below are, concretely, what a human-centered design perspective contributes to OMD practice: not new optimization algorithms or solver improvements, but sustained attention to the socio-technical work that our findings show surrounds every stage of the workflow.
Across the six-stage optimization workflow, practitioners highlighted opportunities for tooling that improve different dimensions of data, dialogue, and decisions to reduce friction, accelerate iteration, and improve reliability.

During \textit{Problem Elicitation}, tools could help clarify requirements, propose alternatives, generate verifiable examples, and support early ideation.
For example, when a stakeholder expresses an underspecified goal like \textit{``this to go up or this to go down''} (P5), an LLM could propose a range of concrete objective functions for the stakeholder to choose from, without having to loop in the OMD immediately.
That said, such suggestions still require mechanisms for error detection and validation.
LLM outputs can appear plausible while being incorrect, and especially non-experts often lack the means to verify them~\citep{zhou2026improving,liao2026chat,zhang2025navigating}.

In \textit{Data Processing}, systems that support early exploration of messy datasets, streamline integration, generate synthetic or placeholder data, and adapt to evolving sources could mitigate major bottlenecks.
As P6 described (see Section~\ref{sec:data_processing}), being able to generate realistic-looking sample data while waiting on the data engineer would have kept that person from becoming a bottleneck.
More broadly, a data-centric perspective suggests that improving the quality, integration, and accessibility of data may matter more than refining models themselves~\citep{jakubik2024data,zha2025data}.
Current workflows and tools rarely prioritize this.

For \textit{Model Development}, tools that map data characteristics to modeling choices, assist in translating requirements into formal models, map problems to canonical forms~\citep{parmentier2022learning}, suggest heuristics or decompositions, and enable quick scenario analysis (e.g., ``what happens to Y if I change X?'') could ease formulation challenges.
LLM-based approaches show promise in converting requirements into mathematical constraints (e.g.,~\citep{lawless2024want}), yet they rarely address other important aspects of OMD practice such as decomposition algorithms~\citep{desaulniers2006column,rahmaniani2017benders,vanderbeck2006generic} or heuristics~\citep{fischetti2010heuristics}.
In terms of using LLMs, optimization has a potential advantage over ML: constraints are generally easier to verify than highly parameterized predictors~\citep{le2025making}, so translation tools may be more inherently transparent.

In \textit{Model Implementation}, interfaces that reduce boilerplate, guide parameter tuning, and support iterative refinement through versioning and pipeline integration would improve adaptability.
More concretely, because implementation is iterative and evolving with new data and stakeholder input, tools that support rapid refinement, lightweight experimentation, and smoother integration with existing data pipelines would help OMDs adapt as problems change.
These opportunities parallel the focus on versioning in MLOps~\citep{shankar2024we}: just as ML engineers track changes in data and pipelines, OMDs could benefit from tools that log evolving problem formulations, constraint modifications, and solver configurations to ensure reliability and reproducibility.

\textit{Model Validation} underscores the need for tools that make optimization models transparent, communicable, and open to critique.
Problem-specific visualizations---such as routing maps or sample schedules---could help stakeholders interpret outputs, while test scenarios probe robustness and expose potential problems in the formulation or data.
These opportunities align with the ``visibility'' dimension in MLOps~\citep{shankar2024we}: optimization tools must surface assumptions, illustrate trade-offs, and make results legible across expertise boundaries.
Importantly, visibility should also foster appropriate reliance~\citep{schemmer2023appropriate,schoeffer2025ai}; for example, through mechanisms that foreground uncertainty, link decisions back to assumptions, or generate comprehensive test cases.
Such features, together with efforts to make optimization algorithms themselves more interpretable~\citep{vcyras2019argumentation,goerigk2023framework,korikov2021counterfactual,kurtz2025counterfactual}, can enable users across expertise levels to critically validate models.

In \textit{Deployment}, structured documentation---akin to model cards~\citep{mitchell2019model} or datasheets in ML~\citep{gebru2021datasheets}---could help offset the time documentation currently takes.
Abstractions that improve portability across solvers and environments would enhance reproducibility and knowledge retention.
Such abstractions, however, risk hiding the solver-specific functionality that experts depend on, and this tension is compounded by reliance on commercial solvers.
Table~\ref{tab:tooling} summarizes the opportunities for tooling and support across workflow steps and their alignment with the three Ds.

Ultimately, studying OMD practice also broadens what CSCW and HCI understand about human-AI collaborative work.
Most of the human-AI interaction literature---especially on trust calibration~\citep{zhang2020effect}, explainability~\citep{schoeffer2024explanations}, and appropriate reliance~\citep{schemmer2023appropriate}---takes as its starting point a system whose internal reasoning is opaque to the people who must act on its outputs; for example, a black-box classifier or an LLM.
The central design problem in this literature, therefore, is how to \textit{produce} legibility where none exists by default.
An optimization model breaks this assumption because it is auditable by construction~\citep{le2025making}, but this auditability does not automatically translate into practical understanding (Section~\ref{sec:ml_comparison}).
This reframes trust, explanation, and appropriate reliance as problems of \textit{operationalizing} an interpretability that already exists: translating formal structure into stakeholder-legible terms, deciding what to surface and when, or building shared vocabulary across formulation and deployment.
For CSCW and HCI scholarship, this suggests a complementary research agenda: not only how to generate explanations for opaque systems, but how to design the socio-technical processes, tools, and interfaces that let auditability-by-construction actually function as a resource for productive collaboration.

\begin{table}[t]
\small
\centering
\caption{Opportunities for tooling and support across workflow stages.}
\label{tab:tooling}
\begin{tabular}{p{0.21\textwidth} p{0.1\textwidth} p{0.6\textwidth}}
\toprule
\textbf{Workflow Step} & \textbf{Three Ds} & \textbf{Opportunities for Tooling and Support} \\
\midrule

Problem Elicitation & Dialogue & Rapidly clarify goals; propose concrete alternatives; generate verifiable examples; support early ideation without repeated back-and-forth \\
\cmidrule{2-3}

Data Processing & Data & Support early exploration of messy datasets; streamline data integration; generate synthetic or placeholder datasets; manage evolving data sources; reduce bottlenecks caused by fragmented pipelines \\
\cmidrule{2-3}

Model Development & Data & Map data characteristics to modeling choices \\
& Decisions & Assist in translating requirements into formal models; map problems to canonical formulations; suggest heuristics or decompositions; enable quick scenario analysis \\
\cmidrule{2-3}

Model Implementation & Decisions & Reduce boilerplate (solver-ready code generation); guide parameter tuning; support iterative refinement (versioning, pipeline integration) \\
\cmidrule{2-3}

Model Validation & Decisions & Surface assumptions and uncertainties; generate robust test scenarios; help stakeholders critically evaluate outputs \\
& Dialogue & Generate visualizations to improve interpretability \\
\cmidrule{2-3}

Deployment & Data & Track assumptions and dependencies \\
& Decisions & Improve portability across solvers and environments  \\
& Dialogue & Document models through structured tools (e.g., model cards) \\
\bottomrule
\end{tabular}
\end{table}

\subsection{Limitations}

Our study offers an important first step toward understanding optimization practice, but it is limited in several ways.
We focused on expert optimization practitioners, rather than non-expert users who might engage with optimization systems more peripherally, leaving questions of accessibility and usability for non-specialists outside the scope of this work.
Relatedly, our analysis emphasized high-level workflows across projects rather than going deeply into any single stage or tool (e.g., data documentation).
Our focus was also limited to model developers themselves, excluding other stakeholders such as managers, domain experts, or end-users of optimization outputs.
This choice allowed us to surface the technical and organizational challenges specific to modeling, but it leaves open how other stakeholders experience or influence the workflow.
In addition, although the OMDs varied in company size, industry, and educational background, we did not systematically analyze subgroup differences.
Finally, our lens was centered on human-centered workflows---how practitioners structure and navigate projects---rather than other approaches to eliciting user input, such as interactive optimization.
\section{Conclusion}

Optimization holds tremendous promise for improving decision-making in domains such as healthcare, logistics, and supply chains.
From the outside, optimization software can appear to function as a ``magical box,'' but our study shows that successful practice depends less on solvers alone and more on the expertise, trade-offs, and collaborations that surround them.
Through interviews with 15 optimization model developers, we surfaced the iterative nature of optimization workflows and highlighted three themes---\textit{data}, \textit{decisions}, and \textit{dialogue}---that cut across all stages.
These themes underscore that optimization is not only a technical exercise but also a socio-technical process grounded in organizational realities.

By characterizing the unique challenges of optimization practice, our work extends CSCW and HCI scholarship beyond prior accounts of ML workflows.
In optimization, data is not simply an input but a pervasive constraint that shapes formulations; decisions are not about prediction accuracy but about satisficing within trade-offs; and dialogue is not incidental but constitutive, enabling co-design, trust, and adoption.
These distinctions carry important design implications: future support tools should \textit{(i)} foreground data exploration and assumption tracking, \textit{(ii)} scaffold satisficing and make trade-offs visible, and \textit{(iii)} embed interactive, transparent mechanisms that sustain dialogue between developers and stakeholders.

Taken together, our findings point toward a new generation of human-centered optimization systems---ones that shift focus from solver-centric pipelines to workflows that are interpretable, collaborative, and aligned with decision-making in practice.

\begin{acks}
MU and CL gratefully acknowledge support from the National Science Foundation (NSF) Award IIS-2233762, the Office of Naval Research (ONR) Awards N000142212825, N000142412306, and N000142312203, IBM, and the Alfred P. Sloan Foundation.
Generative AI tools were used to assist with language editing and manuscript refinement.
All research contributions, analyses, interpretations, and findings were developed by the authors.
\end{acks}

\bibliographystyle{ACM-Reference-Format}
\bibliography{main}

\appendix
\section{Interview Protocol} \label{app:interview_protocol}

\subsection{Welcome}
\paragraph{Introduction}
The goal of this study is to understand the lifecycle of creating and deploying a (mixed-integer) linear programming optimization model to solve real-world problems.
You are asked to participate in one 1-hour session for this study. 
During the study, we will ask you to describe your background and walk us through some of your previous optimization projects.
The goal is to understand your process, how you interface with different stakeholders, and any challenges and pain points.
You do not have to disclose any sensitive or private details about the project, including specific companies, applications, or intellectual property.
For example, instead of saying you built a production planning model to maximize throughput of Coca-Cola at an Atlanta production facility, you can just mention you worked on a model to maximize production of a consumer good at a factory.
Throughout the process, we will be asking you to think aloud to help us understand your thought processes and pain points with your current process for building an optimization model.
Do you have any questions?

\paragraph{Consent to record}
Before we begin, do we have your permission to record the audio of this conversation for the purposes indicated in the consent form that you signed?

\subsection{Background and Experience (5 Minutes)}
We are going to start by learning more about your background and experience with modeling and deploying (mixed-integer) linear programming models in practice.

\begin{itemize}
    \item What is your background in optimization modeling?
    \item What do you and your team do?
    \item Can you ballpark how many (mixed-integer) linear programming optimization models you have helped develop, tweak, or implement during your career? What parts of the modeling process have you worked on?
    \item How long have you been an operations research practitioner? 
    \item Have you worked with other types of optimization models (e.g., queuing systems) or other analytics tools like machine learning or data dashboards?
\end{itemize}
\textit{[If the interview is conducted over Zoom, we will include a short overview of Google Slides and its functionality here.
Specifically, the interviewer will show the user how to add a shape, annotate it with text, and draw arrows between shapes.]}

\subsection{Retrospective Think-Aloud}
For the next part of the interview, I want you to think about one project in which you had to interact with an optimization model.
Try to think about an end-to-end project where you worked on the optimization model from conception to completion if possible. 
\begin{enumerate}
    \item At a high level, what is the pipeline or workflow you followed (starting from the problem to the final model)? While you think about your process, please use the whiteboard and markers in front of you to diagram the different stages of your workflow. Feel free to annotate this diagram with any details about the stage that you think are important to note. \begin{itemize}
        \item How did you learn about the business problem you were trying to solve? Was the problem ambiguous or unclear? How did you resolve any ambiguity / translate the problem into a concrete optimization model? Were there any surprises when you were modeling the program (e.g., business owners with different definitions/constraints than you expected)?
        \item What tools$^*$ did you use for the different stages of this workflow?
        \item Were there any team members or external teams you had to work with? What part of the process did you have to liaise with them?
        \item How did the optimization model relate to other aspects of the broader project?
        \item Was the final optimization model used in practice (by the client or your business)? If not, why not?
    \end{itemize}
    \item How did you evaluate that the model you built was correct? How did you go about debugging it?
    \begin{itemize}
        \item What metrics, if any, did you look at?
        \item Did you have to show the model to any other teams/stakeholders? How did you convince them this was the correct model?
        \item What challenges did you face evaluating the model?
    \end{itemize}
    \item Looking at the entire workflow, what are the biggest challenges you faced? \begin{itemize}
        \item How would you improve the workflow of this project?
        \item Approximately what fraction of the total time of the project did each stage take?
        \item Were there any major blockers that slowed down the project---what stage did they occur in?
        \item Is there anything else about the optimization modeling process you think is important for us to know?
    \end{itemize}
    \item $^*$For every tool mentioned during the lifecycle: \begin{itemize}
        \item How familiar are you with the tool? How frequently do you use it?
        \item Are there any challenges in using the tool?
        \item Is there information or functionality you wish the tool had that it currently does not?
        \item How are you using your time within the tool? 
    \end{itemize}
\end{enumerate}

\subsection{Comparison With Other Tasks}
Now I want you to think about a different project. 
\textit{[If they indicated in the first section that they have done machine learning projects, replace \emph{[PROJECT TYPE]} with \emph{``machine learning''}. Otherwise, if they indicated in the first section that they have done projects with other modeling types (e.g., queuing systems, simulation), replace \emph{[PROJECT TYPE]} with \emph{``other optimization modeling paradigms''}.]}

\paragraph{If additional project type:} You mentioned earlier that you have also worked on projects that involved building a [PROJECT TYPE] model.
\begin{itemize}
    \item Can you think of one such project?
    \item Can you briefly describe the project including the problem you were trying to solve?
    \item Compare the workflow for that project versus the (mixed-integer) linear programming optimization project we just discussed. How does it differ from the workflow you outlined for the last project?
    \item What are the biggest differences between working with a [PROJECT TYPE] model versus a (mixed-integer) linear programming model? 
\end{itemize}

\paragraph{Otherwise:} Now I want you to think about a different (mixed-integer) linear programming project.
Let us compare it to the workflow we just outlined for your first project.
\begin{itemize}
    \item Can you briefly describe the project including the problem you were trying to solve?
    \item Think about your workflow for this project---how does it differ from the workflow you outlined for the last project?
\end{itemize}

\paragraph{Time permitting:}
\begin{itemize}
    \item Can you think of a project that started out as an optimization modeling project but ended up with a different deliverable (e.g., dashboard) or vice versa?
    \item What caused the project to change form? What were the major challenges?
\end{itemize}

\section{Abridged Codebook}
\label{app:codebook}
\sloppy

\subsection*{Problem Elicitation}

\small
\begin{tabular}{p{0.24\linewidth} p{0.32\linewidth} p{0.36\linewidth}}
\toprule
\textbf{Theme} & \textbf{Definition} & \textbf{Illustrative Codes} \\

\midrule

1. Vague initial descriptions &
Early problem statements lack detail and require repeated clarification. &
High-level descriptions; iterative meetings; specific Q\&A; translation from ``English'' to model-ready descriptions. \\[0.5em]

2. Distributed expertise &
Knowledge spans business, engineering, and customer teams, requiring synthesis. &
Fragmented expertise; limited access to key people; tacit assumptions; assimilation time. \\[0.5em]

3. Assessing fit for optimization &
Practitioners evaluate when optimization is appropriate vs.\ simpler solutions. &
Evaluating business value; baseline performance; defining success metrics; alternatives such as dashboards or process changes. \\
\bottomrule

\end{tabular}

\normalsize
\subsection*{Data Processing}

\small
\begin{tabular}{p{0.23\linewidth} p{0.33\linewidth} p{0.38\linewidth}}
\toprule
\textbf{Theme} & \textbf{Definition} & \textbf{Illustrative Codes} \\
\midrule
4. Messy or incomplete data &
Data quality issues require significant cleaning or supplementation. &
Messy/inconsistent data; sparse data; missing probabilities; static vs. dynamic fields; heavy manipulation. \\[0.5em]

5. Slow/difficult data access &
Obtaining data requires coordination and often delays progress. &
Data access delays; unavailable test data; cross-team coordination; bottlenecks. \\[0.5em]

6. Data shapes modeling &
Modeling decisions emerge through exploring available data. &
Iterative data-model shaping; identifying data needs; simulation with simplified assumptions; data documentation; synthetic data. \\
\bottomrule

\end{tabular}

\normalsize

\subsection*{Model Development}

\small
\begin{tabular}{p{0.23\linewidth} p{0.33\linewidth} p{0.38\linewidth}}
\toprule
\textbf{Theme} & \textbf{Definition} & \textbf{Illustrative Codes} \\

\midrule

7. Iterative modeling &
Models evolve with shifting stakeholder needs and constraints. &
Requirement drift; constraint relaxation; multi-month cycles; agile refinement. \\[0.5em]

8. Balancing simplification \& realism &
Practitioners must choose what details to include or abstract. &
Scaffolding; mapping to known templates; choosing details to ignore; satisficing on runtime. \\[0.5em]

9. Literature \& judgment &
Modeling relies on prior expertise and existing operations research methods. &
Identifying problem type; using intuition/coursework; literature search; reusing constraints/variables. \\
\bottomrule

\end{tabular}

\normalsize

\subsection*{Model Implementation}

\small
\begin{tabular}{p{0.23\linewidth} p{0.33\linewidth} p{0.38\linewidth}}
\toprule
\textbf{Theme} & \textbf{Definition} & \textbf{Illustrative Codes} \\

\midrule

10. Iterative prototyping &
Code and constraints are developed incrementally using small tests. &
MVP first; toy problems; constraint-by-constraint testing; pseudocode scaffolding. \\[0.5em]

11. Scalability testing &
Early runs reveal runtime limitations and guide redesign. &
24--48h runtimes; model decompositions; relaxations; business-imposed runtime limits. \\[0.5em]

12. Struggles with optimization solvers &
Solvers are opaque, poorly documented, or difficult to integrate. &
Black-box behavior; poor documentation; parameter tuning; API/portability issues; tool restrictions. \\
\bottomrule

\end{tabular}

\normalsize

\subsection*{Model Validation}

\small
\begin{tabular}{p{0.23\linewidth} p{0.33\linewidth} p{0.38\linewidth}}
\toprule
\textbf{Theme} & \textbf{Definition} & \textbf{Illustrative Codes} \\ 

\midrule

13. Inspection \& critique &
Validation relies on human judgment and stakeholder feedback. &
Sniff tests; manual checks; stakeholder critique; domain-specific visualizations. \\[0.5em]

14. Satisficing on runtime &
Validation balances quality with runtime constraints. &
Accuracy-speed tradeoffs; time budgets; parameter tuning for runtime. \\[0.5em]

15. Combined data and model validation &
Data issues and modeling errors emerge together and require iteration. &
Back-and-forth between data and model; assumption mismatches; scenario testing; data validation tools. \\
\bottomrule

\end{tabular}

\normalsize

\subsection*{Deployment}

\small
\begin{tabular}{p{0.23\linewidth} p{0.33\linewidth} p{0.38\linewidth}}
\toprule
\textbf{Theme} & \textbf{Definition} & \textbf{Illustrative Codes} \\

\midrule

16. Documentation for deployment &
Deployment and maintenance depend on clear documentation and knowledge transfer. &
User/maintainer documentation; handoff; monitoring; explaining models to non-optimization audiences. \\[0.5em]

17. Portability \& integration constraints &
Deployment is shaped by clients' technical infrastructures and constraints. &
Integration with simulation tools; data/format fragmentation; solver portability; IT policy constraints; downstream impacts. \\
\bottomrule

\end{tabular}
\normalsize

\end{document}